\documentclass[11pt]{article}
\usepackage{jheppubmod}
\usepackage{cite}
\usepackage{cancel}
\usepackage{mathrsfs}
\usepackage{enumitem}
\usepackage{xcolor}
\usepackage{caption}  
\usepackage{graphicx} 
\usepackage{float} 
\usepackage{cite}
\usepackage{relsize}
\usepackage{physics}
\usepackage{psfrag}
\usepackage{cancel}
\usepackage{array}
\usepackage{amssymb}
\usepackage{amsmath}
\usepackage[compat=1.1.0]{tikz-feynman}
\usepackage{amsthm}
\usepackage{float}
\usepackage{tikz}
\usetikzlibrary{mindmap,decorations.pathmorphing,backgrounds,positioning,fit}
\usepackage[framemethod=TikZ]{mdframed}

\usetikzlibrary{decorations.markings}
\usepackage{tikz,lipsum,lmodern}
\usepackage[most]{tcolorbox}
\usepackage{hyperref}
\usepackage{xcolor}

\definecolor{twilightlavender}{rgb}{0.54, 0.29, 0.42}
\definecolor{richmaroon}{rgb}{0.69, 0.19, 0.38}
\definecolor{forestgreen(web)}{rgb}{0.13, 0.55, 0.13}
\definecolor{lava}{rgb}{0.81, 0.06, 0.13}
\hypersetup{
	breaklinks,
	colorlinks,
	citecolor=forestgreen(web),
	filecolor=richmaroon,
	linkcolor=lava,
	urlcolor=twilightlavender
}

\tikzset{
pattern size/.store in=\mcSize, 
pattern size = 5pt,
pattern thickness/.store in=\mcThickness, 
pattern thickness = 0.3pt,
pattern radius/.store in=\mcRadius, 
pattern radius = 1pt}
\makeatletter
\pgfutil@ifundefined{pgf@pattern@_ozafp4zcl}{
\pgfdeclarepatternformonly[\mcThickness,\mcSize]{_ozafp4zcl}
{\pgfpointorigin}
{\pgfpoint{\mcSize+\mcThickness}{\mcSize+\mcThickness}}
{\pgfpoint{\mcSize}{\mcSize}}{
\pgfsetcolor{\tikz@pattern@color}
\pgfsetlinewidth{\mcThickness}
\pgfpathmoveto{\pgfpointorigin}
\pgfpathlineto{\pgfpoint{0pt}{0.5*\mcSize}}
\pgfpathlineto{\pgfpoint{\mcSize}{0.5*\mcSize}}
\pgfpathmoveto{\pgfpoint{0.5*\mcSize}{0.5*\mcSize}}
\pgfpathlineto{\pgfpoint{0.5*\mcSize}{\mcSize}}
\pgfpathmoveto{\pgfpoint{0pt}{\mcSize}}
\pgfpathlineto{\pgfpoint{\mcSize}{\mcSize}}
\pgfusepath{stroke}}}
\makeatother

\tikzset{
pattern size/.store in=\mcSize, 
pattern size = 5pt,
pattern thickness/.store in=\mcThickness, 
pattern thickness = 0.3pt,
pattern radius/.store in=\mcRadius, 
pattern radius = 1pt}
\makeatletter
\pgfutil@ifundefined{pgf@pattern@name@_nerwo1frk}{
\pgfdeclarepatternformonly[\mcThickness,\mcSize]{_nerwo1frk}
{\pgfqpoint{0pt}{-\mcThickness}}
{\pgfpoint{\mcSize}{\mcSize}}
{\pgfpoint{\mcSize}{\mcSize}}
{
\pgfsetcolor{\tikz@pattern@color}
\pgfsetlinewidth{\mcThickness}
\pgfpathmoveto{\pgfqpoint{0pt}{\mcSize}}
\pgfpathlineto{\pgfpoint{\mcSize+\mcThickness}{-\mcThickness}}
\pgfusepath{stroke}
}}
\makeatother
\def\ee{\end{equation}}
\def\be{\begin{equation}}
\def\bea{\begin{eqnarray}}
\def\eea{\end{eqnarray}}
\newcommand{\beq}{\begin{eqnarray}}
\newcommand{\eqq}{\end{eqnarray}}
 \newcommand{\badat}{\begin{alignedat}}
 \newcommand{\eadat}{\end{alignedat}}

\newcommand{\eal}[1]{\be \begin{aligned} #1 \end{aligned}\end{equation}} 
\newcommand{\eqn}[1]{\be #1 \end{equation}} 
\newcommand{\eqa}[1]{\bea  #1\end{eqnarray}}

\renewcommand{\d}{\mathrm{d}}

\newcommand{\eg}{{\it e.g.,\ }}

\long\def\new#1\endnew{{\bf #1}}		
\long\def\del#1\enddel{}

\def\del{\partial}

\usepackage{color}

\newcommand{\pink}[1]{\textcolor{\pink}{#1}}

\definecolor{dblue}{rgb}{0.2,0.50,0.80}

\title{Spectral representation in Klein space: simplifying celestial leaf amplitudes}
\affiliation[\blacklozenge]{International Centre for Theoretical Sciences-TIFR, Shivakote, Hesaraghatta Hobli, Bengaluru North
560089, India.}
\affiliation[\lozenge]{School of Basic Sciences, Indian Institute of Technology Bhubaneswar, Jatni, Khurda, Odisha, 752050, India.}
\usepackage{orcidlink}
\author[\blacklozenge,\orcidlink{0000-0002-4535-3198}]{Sarthak Duary,}\emailAdd{sarthak.duary@icts.res.in}
\author[\lozenge,\orcidlink{0009-0008-2963-2497}]{and Sourav Maji}\emailAdd{sm89@iitbbs.ac.in}
\abstract{In this paper, we explore the spectral representation in Klein space, which is the split $(2,2)$ signature flat spacetime. The Klein space can be foliated into Lorentzian $\mathrm{AdS}_3 /\mathbb{Z}$ slices, and its identity resolution has continuous and discrete parts. We calculate the identity resolution and the Plancherel measure in these slices. Using the foliation of Klein space into the slices, the identity resolution, and the Plancherel measure in each slice, we compute the spectral representation of the massive bulk-to-bulk propagator in Klein space. It can be expressed as the sum of the product of two massive (or tachyonic) conformal primary wavefunctions, with both continuous and discrete parts, and sharing a common boundary coordinate. An interesting point in Klein space is that, since the identity resolution has discrete and continuous parts, a new type of conformal primary wavefunction naturally arises for the massive (or tachyonic) case. For the conformal primary wavefunctions, both the discrete and continuous parts involve integrating over the common boundary coordinate and the real (or imaginary) mass. The conformal dimension is summed in the discrete part, whereas it is integrated in the continuous part. The spectral representation in Klein space is a computational tool to derive conformal block expansions for celestial amplitudes in Klein space and its building blocks, called celestial leaf amplitudes, by integrating the particle interaction vertex over a single slice of foliation.}
\begin{document}

\maketitle
\section{Introduction}
\label{intro}

The celestial holography initiative seeks to extend the holographic principle to spacetimes with a vanishing cosmological constant, drawing inspiration from the celebrated AdS/CFT duality \cite{Maldacena:1997re,Witten:1998qj, Aharony:1999ti}. Unlike the top-down methodology of the AdS/CFT duality, celestial holography has predominantly evolved through a bottom-up strategy. The formidable challenge of constructing dual pairs from a top-down standpoint has largely remained unexplored, with only a handful of recent proposals, primarily focused on the self-dual sector \eg see \cite{costello2023top, costello2023burns}. The objective in the celestial holography initiative is twofold: firstly, to exploit this duality to probe non-perturbative aspects of quantum gravity, and secondly, to lay the groundwork for celestial conformal field theories (CCFTs). Alternatively, it can be viewed as a CFT-centric approach to exploring amplitudes and earlier $\mathcal{S}$-matrix programs \cite{Eden:1966dnq,Elvang:2013cua,Henn:2014yza}. By fostering a deeper understanding of consistent CCFTs, it becomes viable to apply CFT-bootstrap techniques. In the celestial holography initiative, the principal emphasis is placed on the $\mathcal{S}$-matrix. What distinguishes this approach is the methodical development of a framework where symmetries are not merely fundamental but also infinite-dimensional in nature.
%.....\texttt{To be added more}
%The celestial holography program seeks to apply the holographic principle to spacetimes with vanishing cosmological constant. This is motivated by the AdS/CFT duality \cite{Maldacena:1997re,Witten:1998qj, Aharony:1999ti}. 
%Unlike the AdS/CFT duality, celestial holography has mainly developed through a bottom-up approach. The difficult task of creating dual pairs from the top-down perspective has largely been unaddressed, save for a handful of recent suggestions mainly centered on the self-dual sector, \eg see\cite{costello2023top, costello2023burns}. The aim is to utilize this duality not only to address non-perturbative aspects of quantum gravity but also to establish a framework for celestial conformal field theories (CCFTs). Alternatively, it can be seen as a CFT-driven approach to the study of amplitudes and earlier $\mathcal{S}$-matrix programs\cite{Eden:1966dnq,Elvang:2013cua,Henn:2014yza}. By developing an intrinsic understanding of consistent CCFTs, it becomes useful to employ CFT-bootstrap techniques. In the celestial holography program, the primary focus lies on the $\mathcal{S}$-matrix. What sets this approach apart is the systematic construction of a framework where symmetries are not only fundamental but infinite dimensional. 

This branch of research is spurred by Strominger's astute observation \cite{Strominger:2013jfa,Strominger:2013lka} regarding soft theorems in quantum field theory \cite{Weinberg:1965nx}, which are reflections of Ward identities associated with asymptotic symmetries \cite{Bondi:1962px,Sachs:1962wk,Sachs:1962zza}. Notably, these asymptotic symmetry generators resemble currents residing on the celestial sphere, albeit in two lower dimensions. Also, the subleading soft graviton theorem \cite{Cachazo:2014fwa}, linked with an asymptotic Virasoro symmetry \cite{Kapec:2014opa}, gives rise to a stress tensor. Nonetheless, to conform to the standard form of the conformal Ward identity, a transition from momentum to boost eigenstates \cite{Pasterski:2016qvg,Pasterski:2017kqt,Pasterski:2017ylz} is required.

Celestial holography proposes a correspondence between a $(d+2)$-dimensional quantum field theory or quantum gravity in flat space and a $d$-dimensional CCFT on the celestial sphere at null infinity. This suggests that when the $\mathcal{S}$-matrix in the higher-dimensional bulk theory is expressed in terms of conformal primary basis, exhibiting $d$-dimensional conformal symmetry (or equivalently, $(d+2)$-dimensional Lorentz symmetry), they correspond to correlation functions in the boundary theory \cite{Pasterski:2016qvg,Pasterski:2017kqt}. The subleading soft graviton theorem \cite{Cachazo:2014fwa} implies that observables within any consistent $4d$ quantum theory of gravity in asymptotically flat-space remain invariant under transformations by the local $2d$ conformal group acting on the celestial sphere \cite{Kapec:2014opa}. This introduces the powerful framework of $2d$ CFT to address the intricate challenges of $4d$ quantum gravity. Notably, the scattering amplitudes of $4d$ gravity adhere to the same constraints as those governing $2d$ CFT correlators. However, applying the toolkit of CFT to constrain quantum gravity scattering amplitudes has encountered hurdles, primarily due to the distributional form \footnote{Such forms are unfamiliar within the realm of $2d$ CFT studies. Particularly, momentum space delta functions may appear as complex contact interactions in the $2d$ correlators, making them challenging to analyze.} taken by these amplitudes, despite their full conformal invariance.

Now, alongside conformal symmetry, another essential criterion for a CFT is locality. Locality, regarding correlation functions, asserts that singularities occur only when operators coincide, and these singularities are characterized by operator product expansions (OPEs). Notably, studying the celestial OPEs of gluons and gravitons reveals that singularities at coinciding coordinates on the celestial sphere correspond to collinear singularities in scattering amplitudes \cite{Pate:2019lpp}. To examine the singularity structure of celestial amplitudes, it is useful to fully leverage conformal symmetry by expanding them in terms of conformal blocks. Prior works have shown the conformal block expansions of celestial four-point massless scalar amplitudes in \cite{Lam:2017ofc,Atanasov:2021cje,Chang:2021wvv}, utilizing the completeness relation of conformal partial waves. However, extending this approach to higher-point functions presents challenges due to integrals over the space of conformal cross ratios. To tackle this challenge, one can employ AdS holography, where a computational technique for computing AdS Witten diagrams and their conformal block expansions involves the spectral representation of AdS bulk-to-bulk propagators \cite{Penedones:2010ue, Costa:2014kfa, Meltzer:2019nbs}.

In \cite{Chang:2023ttm}, Chang, Liu, and Ma (CLM) introduce a spectral representation prescription for celestial amplitudes. 
This CLM spectral representation prescription involves cutting internal lines of Feynman diagrams in Minkowski space. The spectral representations for celestial amplitudes have also been studied in \cite{Melton:2021kkz}. 

In the framework of hyperbolic foliations of Minkowski space \eg see \cite{deBoer:2003vf}, Minkowski space is foliated by regions having past or future light cones (Euclidean AdS slices) and regions outside these cones (dS slices). Using the CLM prescription, the Euclidean AdS and dS spectral representations are applied separately to the parts of the bulk-to-bulk propagator corresponding to the Euclidean AdS and dS slices. Combining these spectral representations yields the final spectral representation for the bulk-to-bulk propagator in Minkowski space. To elaborate, in the CLM prescription the bulk-to-bulk propagators for the internal lines can be represented as a product of two boundary-to-bulk propagators, where a shared boundary coordinate is integrated over the celestial sphere. By employing the CLM spectral representation prescription, one can carry out computations for the conformal block expansions of celestial amplitudes. 

Now, analytic continuation from $(1,3)$ Lorentzian to $(0,4)$ Euclidean signature has proven valuable in addressing scattering problems. Recently, the analytic continuation from $(1,3)$ Lorentzian to $(2,2)$ Kleinian signature, i.e. for Klein space has arisen as a tool for following reasons.

\begin{enumerate}[label=\roman*.]
\item On-shell techniques and spinor-helicity variables, has resulted in formulations much simpler than those derived from the conventional Feynman diagram approach, \eg see \cite{Elvang:2013cua}. In many cases, spinor variables in conjugate representations are handled as independent entities, either explicitly or implicitly, particularly in the realm of holomorphic soft limits, which play a significant role in governing their structure. \cite{Cachazo:2014fwa}. This approach is tantamount to operating within a $(2,2)$ signature, where the Lorentz group extends into two distinct copies, one self-dual and the other anti-self-dual, of $\mathrm{SL}(2,\mathbb{R})$. 
\item The scattering amplitudes are governed by the on-shell three-point function \cite{Britto:2005fq,Cachazo:2004kj,Arkani-Hamed:2008bsc}. Yet, in the Euclidean space, nothing can achieve an on-shell state (save for tachyons!). Meanwhile, in Minkowski space, the kinematics make massless three-point amplitudes zero. Therefore, describing massless scattering requires the more complex four-point interactions. In contrast, in Klein space, massless three-point amplitudes are usually nonzero and serve as a valuable characterization of the theoretical framework.
\end{enumerate}

In this paper, we develop the spectral representation in Klein space. Klein space can be divided into two sets of hyperbolic slices in the timelike and spacelike wedges. These slices consist of Lorentzian AdS$_3/\mathbb{Z}$ spacetimes, with their boundaries forming copies of the celestial torus. An AdS$_3/\mathbb{Z}$-Witten diagram is associated with each slice. %giving rise to a conformally invariant leaf amplitude. 
 We use the foliation of Klein space into slices of $\mathrm{AdS}_3/\mathbb{Z}$ and resolving identities on these slices to get the spectral representation of the massive bulk-to-bulk propagator in Klein space. The spectral representation in Klein space is a computational tool for deriving conformal block expansions for celestial amplitudes in Klein space along with its building blocks known as celestial leaf amplitudes. This involves integrating the particle interaction vertex across a single slice of foliation. \footnote{Spacetime translation invariance imposes several relations among already tightly constrained conformally invariant correlators, leading to instances where these constraints can only be satisfied through distributional expressions. Nonetheless, several successful attempts have been made to construct the conformally invariant celestial amplitudes by expanding around backgrounds that are not translation invariant \cite{Costello:2022wso,Fan:2022elem,Casali:2022fro,Gonzo:2022tjm,Stieberger:2023fju,Costello:2022jpg,Stieberger:2022zyk,Melton:2022fsf,Bittleston:2023bzp,Costello:2023hmi,Adamo:2023zeh,Ball:2023ukj}. Due to the integration of particle interaction vertices over a single slice of foliation within a hyperbolic spacetime foliation, celestial leaf amplitudes exhibit conformal invariance but lack translation invariance. This distinction imposes a distributional structure on celestial amplitudes, whereas leaf amplitudes are smooth.} Our primary goal in utilizing the spectral representation in Klein space to set up the formalism to explore the conformal block expansions of celestial amplitudes in Klein space.

Now, we give the structure, concept map, and summary of the paper.
\paragraph{Structure, concept map, and summary.} 
As a first  step, we start with the kinematics of Klein space in section \ref{kkin}. 
We study the Klein space from Lorentzian flat-space via analytic continuation in subsection \ref{anacon}. We study the hyperbolic foliation of Klein space by Lorentzian $\mathrm{AdS}_3/\mathbb{Z}$ slices in subsection \ref{hfol} and discuss celestial amplitudes in Klein space in subsection \ref{ctor}. We explicitly show that the identity resolution on $\mathrm{AdS}_3/\mathbb{Z}$ consists of both continuous and discrete parts in section \ref{dc}. We calculate the Plancherel measure in subsection \ref{planmeas} and identity resolution in $\mathrm{AdS}_3/\mathbb{Z}$ in subsection \ref{ires}, and establish a connection to the Plancherel measure of the CLM prescription for Euclidean AdS space and dS spacetime \cite{Chang:2023ttm}, which is consistency check for our prescription.
Finally, we use the Klein space foliation into $\mathrm{AdS}_3/\mathbb{Z}$ slices and the identity resolution on these slices to derive the spectral representation of the massive bulk-to-bulk propagator in Klein space in section \ref{sr}. In this paper, we aim to keep the content concise and avoid technical complexities. However, for intrepid readers, we give a comprehensive exposition of the details in the appendix. In appendix \ref{appA}, we give some mathematical preliminaries. In appendix \ref{sleqnn}, we study the laplacian in $\mathrm{AdS}_3/\mathbb{Z}$ global coordinate, and the Sturm-Liouville equation. We also give some remark on the Sturm-Liouville equation about the dependence of dimension in arbitrary dimension. In the appendix \ref{alder}, we give an alternative derivation of the Plancherel measure as a special case: case $p=q=2$.
To simplify things for the reader, we give a structure, concept map, and summary of the paper.

\begin{figure}[htbp]
    \centering
    \includegraphics[width=1\textwidth]{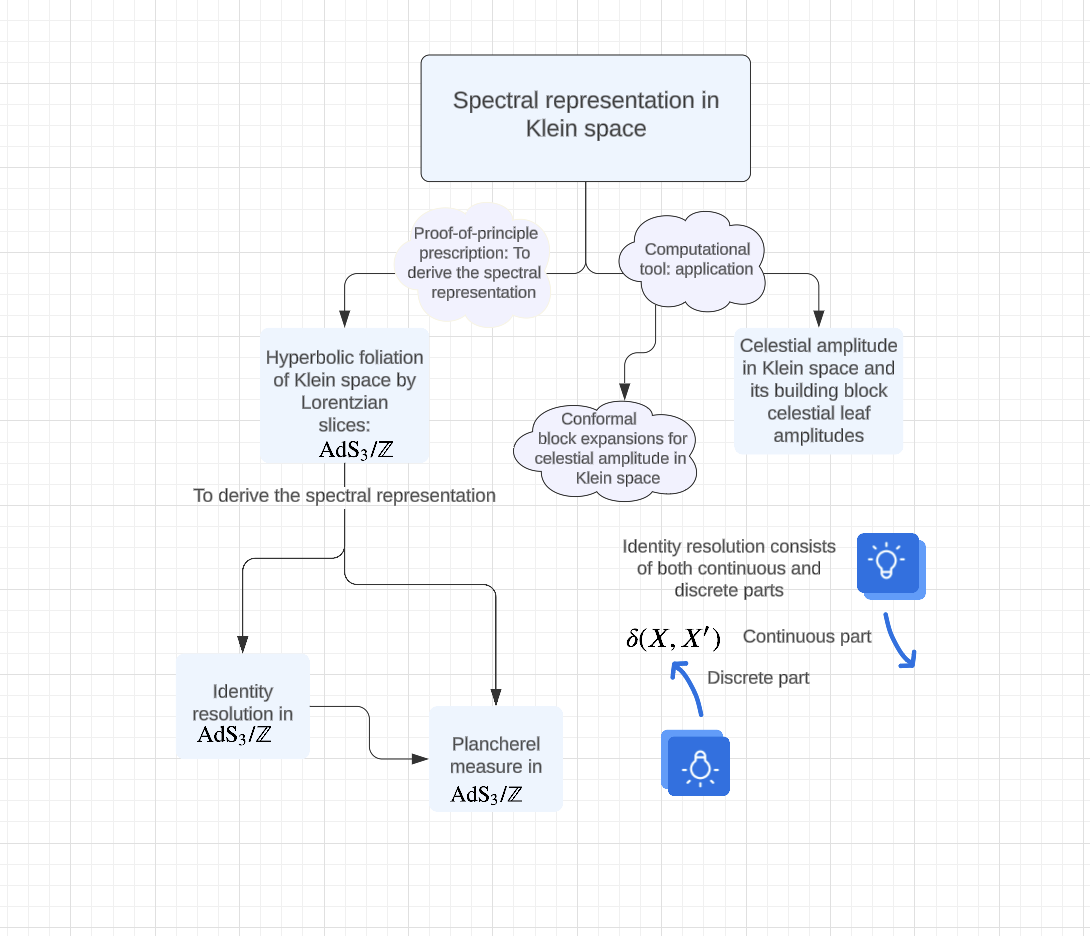}
    \caption{Concept map}
\label{conmap}    
\end{figure}

We give the concept map in the figure \ref{conmap}.
Using the resolution of identity in $\mathrm{AdS}_3/\mathbb{Z}$, which includes both continuous and discrete parts, we compute the spectral representation of the massive bulk-to-bulk propagator in Klein space. An interesting point to note is that as the identity resolution has discrete and continuous parts, it gives rise to a new type of conformal primary wavefunction in the context of the massive (or tachyonic) case. This spectral representation can be described as the product of two massive (or tachyonic) conformal primary wavefunctions with real (or imaginary) mass, each having continuous and discrete parts with conformal dimensions $\Delta$ and $2-\Delta$, with a common boundary point. For both the discrete and continuous parts we have integration over the common boundary coordinate and the real (or imaginary) mass. For the discrete part, the conformal dimension $\Delta$ is summed, while for the continuous part, it is integrated.

\section{Klein space kinematics}
\label{kkin}
In this section \ref{kkin}, we study hyperbolic foliation of Klein space, celestial amplitude in Klein space, and its building blocks called celestial leaf amplitude. %This celestial leaf amplitude construction is most naturally formulated in the Klein space $\mathbb{R}^{2,2}$, i.e. the flat-space with signature $(2, 2)$.
For the details to the relevant Kleinian geometry utilized in our approach \eg see \cite{Melton:2024jyq, Melton:2023bjw, Melton:2023hiq, Casali:2022fro, Atanasov:2021oyu, Bhattacharjee:2021mdc}.

\subsection{Klein space from Lorentzian flat-space via analytic continuation}
\label{anacon}
The Lorentzian flat-space metric is
\begin{equation}
    \d s_{\textrm{Mink}}^2=-(\mathrm{d} X^0)^2+(\mathrm{d}X^1)^2 + (\mathrm{d}X^2)^2 + (\mathrm{d} X^3)^2.
\end{equation}
The analytic continuation we have 
\begin{align}
X^1\to i X^1. \label{eq:anac}
\end{align}
After applying the analytic continuation of eq.\eqref{eq:anac} gives
\begin{align}
    \d s^2_{\text{Klein}}=&-(\mathrm{d} X^0)^2 -(\mathrm{d}X^1)^2 + (\mathrm{d}X^2)^2 + (\mathrm{d} X^3)^2 \label{eq_flatmetricconf1}\\
    =& - \d\varrho^2 -\varrho^2 \d\zeta^2 +\d \mathfrak{R}^2 + \mathfrak{R}^2 \d \xi^2. \label{eq_flatmetricconf}
\end{align}
\begin{figure}[H]
	\centering
	\tikzset{every picture/.style={line width=0.75pt}} %set default line width to 0.75pt        
\begin{tikzpicture}[x=0.8pt,y=0.8pt,yscale=-0.8,xscale=0.8]
%uncomment if require: \path (0,466); %set diagram left start at 0, and has height of 466

%Shape: Right Triangle [id:dp9202705119849155] 
\draw  [color={rgb, 255:red, 0; green, 0; blue, 0 }  ,draw opacity=1 ][line width=1.5]  (208,95.73) -- (459,346.73) -- (208,346.73) -- cycle ;
%Curve Lines [id:da05746797478013388] 
\draw    (208,95.73) .. controls (242,196.73) and (261,262.73) .. (259,346.73) ;
%Curve Lines [id:da26238687617906686] 
\draw    (208,95.73) .. controls (268.33,196.73) and (310.33,261.4) .. (308.33,345.4) ;
%Curve Lines [id:da5976223747130731] 
\draw    (208,95.73) .. controls (286.33,195.4) and (359.67,260.73) .. (357.67,344.73) ;
%Curve Lines [id:da15679335033559583] 
\draw  [dash pattern={on 4.5pt off 4.5pt}]  (458.86,347) .. controls (359.3,308.99) and (294.78,297.37) .. (210.77,296.01) ;
%Curve Lines [id:da20669299560507515] 
\draw  [dash pattern={on 4.5pt off 4.5pt}]  (459,346.73) .. controls (360.49,282.41) and (292.75,246.8) .. (208.74,245.44) ;
%Curve Lines [id:da040797363377256035] 
\draw  [dash pattern={on 4.5pt off 4.5pt}]  (458.86,347) .. controls (362.41,264.75) and (291.68,197.69) .. (207.67,196.33) ;
%Straight Lines [id:da5812307251015338] 
\draw [color={rgb, 255:red, 74; green, 144; blue, 226 }  ,draw opacity=1 ][line width=2.25]    (207.61,346.91) -- (207.61,267.2) ;
\draw [shift={(207.61,262.2)}, rotate = 450] [fill={rgb, 255:red, 74; green, 144; blue, 226}  ,fill opacity=1 ][line width=0.08]  [draw opacity=0] (14.29,-6.86) -- (0,0) -- (14.29,6.86) -- cycle    ;
%Straight Lines [id:da8054047670313162] 
\draw [color={rgb, 255:red, 74; green, 144; blue, 226 }  ,draw opacity=1 ][line width=2.25]    (207.61,346.91) -- (287.32,346.91) ;
\draw [shift={(292.32,346.91)}, rotate = 180] [fill={rgb, 255:red, 74; green, 144; blue, 226}  ,fill opacity=1 ][line width=0.08]  [draw opacity=0] (14.29,-6.86) -- (0,0) -- (14.29,6.86) -- cycle    ;
%Straight Lines [id:da8640571871909553] 
\draw    (398,100) -- (457,100) ;
%Straight Lines [id:da49847496081359877] 
\draw  [dash pattern={on 4.5pt off 4.5pt}]  (399,133) -- (458,133) ;

% Text Node
\draw (210.45,71.6) node  [font=\LARGE,color={rgb, 255:red, 0; green, 0; blue, 0 }  ,opacity=1 ]  {$i'$};
% Text Node
\draw (342.35,191.2) node  [font=\LARGE,rotate=-359]  {$\mathcal{I}$};
% Text Node
\draw (479.78,348.67) node  [font=\LARGE,color={rgb, 255:red, 0; green, 0; blue, 0 }  ,opacity=1 ]  {$i^{0}$};
% Text Node
\draw (190.47,276.87) node  [font=\LARGE,color={rgb, 255:red, 74; green, 144; blue, 226}  ,opacity=1 ] [align=left] {$\displaystyle \varrho $};
% Text Node
\draw (283.8,366.2) node  [font=\LARGE,color={rgb, 255:red, 74; green, 144; blue, 226}  ,opacity=1 ] [align=left] {$\displaystyle \mathfrak{R}$};
% Text Node
\draw (501.57,97) node  [font=\Large] [align=left] {$\displaystyle \mathfrak{R}$ const.};
% Text Node
\draw (502.32,131) node  [font=\Large] [align=left] {$\displaystyle \varrho $ const.};
\end{tikzpicture}	
	\caption{The Penrose diagram depicting Klein space has null lines positioned at a 45$^\circ$ angle, along with lines representing constant values of \(\varrho\) and \(\mathfrak{R}\). It also includes labels for timelike (\(i'\)), spacelike (\(i^0\)), and null infinity (\(\mathcal{I}\)).}
	\label{fig:penrose-diagram}
\end{figure}
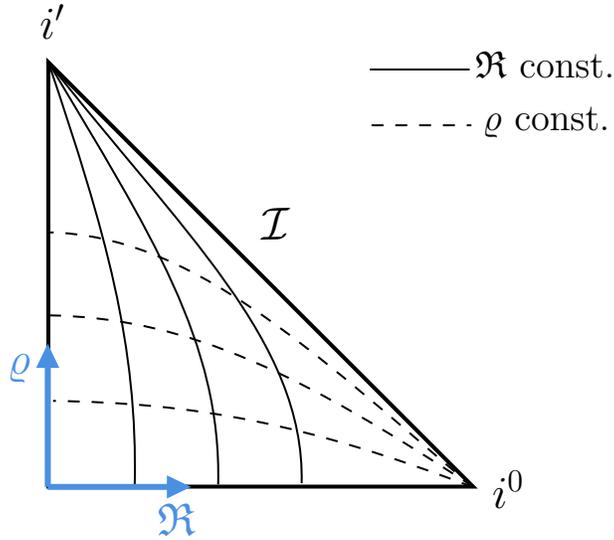

Here, $\varrho, \mathfrak{R} \geq 0$ and $\zeta, \xi \in [0, 2\pi)$, and we have
\begin{equation}
\begin{split}
X^0 &= \varrho \cos \zeta \\
X^1 &= \varrho \sin \zeta \\
X^2 &= \mathfrak{R} \cos \xi \\
X^3 &= \mathfrak{R} \sin \xi.
\end{split}
\end{equation}
The Penrose diagram on the \(\mathfrak{R}\)-\(\varrho\) plane is obtained through conformal compactification of eq.\eqref{eq_flatmetricconf}, as illustrated in figure \ref{fig:penrose-diagram}. Each point within the diagram represents a Lorentzian torus formed by the two \(U(1)\) cycles associated with \(\zeta\) and \(\xi\). 

The conformal metric at null infinity ($\mathcal{I}$) is constructed as a product of a square Lorentzian torus and a null interval. In this configuration, the spacelike cycle of the torus degenerates along the timelike line $\mathfrak{R}=0$, meaning it collapses or shrinks to zero size at this line. Similarly, the timelike cycle degenerates along the spacelike line $\rho=0$, collapsing at this point. To maintain the integrity of the cycles and prevent them from degenerating within the interior of the space, the condition $\mathfrak{R}, \rho \geq 0$ is imposed in eq.\eqref{eq_flatmetricconf}. This restriction ensures that the cycles remain non-degenerate except precisely at the boundary lines, preserving the overall structure of the conformal metric throughout the rest of the space.

\subsection{Hyperbolic foliation of Klein space}
\label{hfol}
In this subsection \ref{hfol}, we study the hyperbolic foliation of Klein space. For the discussion on the hyperbolic foliation of Klein space, we refer to \cite{Melton:2023bjw}.

Let us begin by examining the structure of Klein space, which is a flat-space with two timelike and two spacelike dimensions. It is organized into slices known as Lorentzian $\text{AdS}_3/\mathbb{Z}$ slices, or `leaves'. The boundary of this space, where the dual CCFT resides, consists of celestial tori. These tori represent parameterizations of null momenta within the Klein space. 

The metric of Klein space $\mathbb{K} = \mathbb{R}^{2,2}$, i.e. Minkowski space to $(2, 2)$ split signature with coordinates $X^\mu$ is given by 
\begin{equation}
\mathrm{d} s_{\text{Klein}}^2 = -(\mathrm{d} X^0)^2 -(\mathrm{d}X^1)^2 + (\mathrm{d}X^2)^2 + (\mathrm{d} X^3)^2.
\end{equation}
Klein space divides into a timelike wedge denoted as $\mathcal{W}^T$ where $X^2<0$, and a spacelike wedge denoted as $\mathcal{W}^S$ where $X^2>0$
\begin{equation}
\begin{split}
\mathcal{W}^T~:~ X^\mu = \tau\hat x_+^\mu~,~\hat x_+^2=-1,\\
\mathcal{W}^S~:~ X^\mu = \tau\hat x_-^\mu~,~\hat x_-^2=+1.
\end{split}
\end{equation}
Here, $\tau\in(0,\infty)$ is the magnitude of $X^\mu$.

\begin{figure}[H]
	\centering
	\tikzset{every picture/.style={line width=0.75pt}} %set default line width to 0.75pt        

\begin{tikzpicture}[x=0.75pt,y=0.75pt,yscale=-1,xscale=1]
	%uncomment if require: \path (0,300); %set diagram left start at 0, and has height of 300
	\draw  [pattern=_ozafp4zcl,pattern size=6pt,pattern thickness=0.75pt,pattern radius=0pt, pattern color={rgb, 255:red, 80; green, 227; blue, 194}] (210,140) -- (100,250) -- (100,130) -- (100,30) -- (100,30) -- cycle ;
	%Shape: Polygon [id:ds9029030036980761] 
	\draw  [pattern=_nerwo1frk,pattern size=6pt,pattern thickness=0.75pt,pattern radius=0pt, pattern color={rgb, 255:red, 144; green, 19; blue, 254}] (210,140) -- (320,250) -- (100,250) -- (210,140) -- (210,140) -- cycle ;
	%Shape: Right Triangle [id:dp48105542387732736] 
	\draw   (100,30) -- (320,250) -- (100,250) -- cycle ;
	%Straight Lines [id:da3227447501527644] 
	\draw  [dash pattern={on 0.84pt off 2.51pt}]  (100,250) -- (210,140) ;
	%Curve Lines [id:da1577778616294202] 
	\draw    (100,210) .. controls (135.14,203.65) and (177.14,167.65) .. (210,140) ;
	%Curve Lines [id:da01214480998814882] 
	\draw    (100,180) .. controls (136.14,185.65) and (176.14,165.65) .. (210,140) ;
	%Curve Lines [id:da9812793985793039] 
	\draw    (100,110) .. controls (134.14,109.65) and (176.14,117.65) .. (210,140) ;
	%Curve Lines [id:da9490483303784552] 
	\draw    (100,70) .. controls (134.14,69.65) and (176.14,117.65) .. (210,140) ;
	%Curve Lines [id:da8877140932372717] 
	\draw    (140,250) .. controls (140.14,233.65) and (175.14,170.65) .. (210,140) ;
	%Curve Lines [id:da874226903107285] 
	\draw    (170,250) .. controls (168.14,233.65) and (183.14,171.65) .. (210,140) ;
	%Curve Lines [id:da7690433309872561] 
	\draw    (250,250) .. controls (251.14,228.65) and (239.14,170.65) .. (210,140) ;
	%Curve Lines [id:da9863017982325144] 
	\draw    (280,250) .. controls (281.14,228.65) and (239.14,170.65) .. (210,140) ;
	%Shape: Polygon [id:ds6232442003310572] 

	% Text Node
	\draw (109,138) node [anchor=north west][inner sep=0.75pt]    {$\mathcal{W}^{T}$};
	% Text Node
	\draw (190,198) node [anchor=north west][inner sep=0.75pt]    {$\mathcal{W}^{S}$};
	% Text Node
	\draw (214,109) node [anchor=north west][inner sep=0.75pt]    {$\mathcal{I}$};
\end{tikzpicture}
	\caption{A Penrose diagram illustrating the hyperbolic foliation of Klein space, having both its timelike ($\mathcal{W}^T$) and spacelike ($\mathcal{W}^S$) wedges. It includes labels for null infinity (\(\mathcal{I}\)).}
	\label{fig:toric-penrose-diagram}
\end{figure}
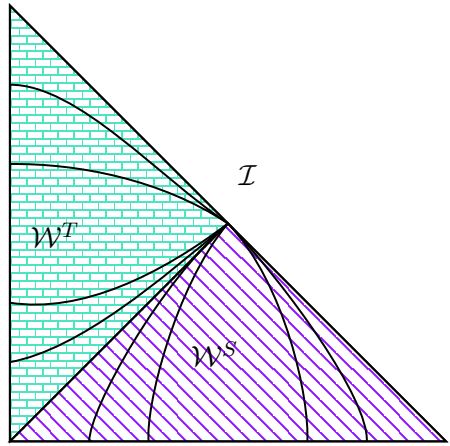

In any given region, surfaces of constant $\tau$ are diffeomorphic to $\mathrm{AdS}_3/\mathbb{Z}$, representing Lorentzian AdS space with periodically identified time. When combined, they form a hyperbolic foliation of Klein space, analogous to the hyperbolic foliation of Minkowski space \eg see \cite{deBoer:2003vf}. The timelike surfaces defined by $\hat x^2_+=-1$, which form the foliation of the timelike wedge $\mathcal{W}^T$, can be described using standard global coordinates
\begin{equation}
\hat x_+^0+i \hat x_+^1 = e^{i \psi}\cosh\rho~,~\hat x_+^2+i\hat x_+^3 = e^{i \phi}\sinh\rho.
\end{equation}
Here, $\rho$ is the radial coordinate on $\mathrm{AdS}_3/\mathbb{Z}$ and ranges from $0$ to $\infty$ , i.e. $\rho \in (0, \infty)$. The coordinates $\psi$ and $\phi$ are periodic, with $\psi$ having a periodicity of $\psi \sim \psi + 2\pi$ along the timelike cycle, and $\phi$ having a periodicity of $\phi \sim \phi + 2\pi$ along the spacelike cycle.
%Here, $\rho$ lies in the interval $(0, \infty)$, i.e. $\rho \in (0, \infty)$ and serves as the radial coordinate on $\mathrm{AdS}_3/\mathbb{Z}$, while $\psi$ (with periodicity $\psi \sim \psi + 2\pi$) and $\phi$ (with periodicity $\phi \sim \phi + 2\pi$) are the periodic coordinates along the timelike and spacelike cycles, respectively. 
The metric is
\begin{equation}
\begin{split}
\mathrm{d} s_{\text{Klein}}^2&=-\mathrm{d} \tau^2 +\tau^2\big(-\cosh^2\rho \mathrm{d} \psi^2+\sinh^2\rho \mathrm{d} \phi^2+\mathrm{d} \rho^2\big)\\
&=-\mathrm{d} \tau^2 +\tau^2 \mathrm{d} s_3^2,
\end{split}
\end{equation}
where the bracketed part $\mathrm{d} s_3^2$ corresponds to the unit metric on $\mathrm{AdS}_3/\mathbb{Z}$. 
The unit metric on $\mathrm{AdS}_3/\mathbb{Z}$ is given by
\begin{equation}
\mathrm{d} s_3^2=-\cosh^2\rho \mathrm{d} \psi^2+\sinh^2\rho \mathrm{d} \phi^2+\mathrm{d} \rho^2.
\end{equation}
The $\mathbb{Z}$ acts as the time-like quotient $\psi \sim \psi+2\pi$.
The presence of the $\mathbb{Z}$-quotient introduces periodicity to the conventional $\mathrm{AdS}_3$ cylinder in this context. Similarly, the spacelike wedge $\mathcal{W}^S$ (defined by $\hat x_-^2=+1$) can be represented using analogous coordinates, albeit with the roles of spacelike and timelike cycles interchanged 
\begin{equation}
	\hat x_+^0+i \hat x_+^1 = e^{i \psi}\sinh\rho~,~\hat x_+^2+i\hat x_+^3 = e^{i \phi}\cosh\rho.
\end{equation}
The metric is 
\begin{equation}
\begin{split}
\mathrm{d} s_{\text{Klein}}^2&=-\mathrm{d} \tau^2 -\tau^2\big(-\cosh^2\rho \mathrm{d} \phi^2+\sinh^2\rho \mathrm{d} \psi^2+\mathrm{d} \rho^2\big)\\
&=-\mathrm{d} \tau^2 -\tau^2 \mathrm{d} s_3^2.
\end{split}
\end{equation}
A Penrose diagram representing the hyperbolic foliation of Klein space in both its timelike ($\mathcal{W}^T$) and spacelike ($\mathcal{W}^S$) wedges is given in the figure \ref{fig:toric-penrose-diagram}. 
% TODO: \usepackage{graphicx} required
The constant-$\tau$ slices, both timelike and spacelike, exhibit the geometry of $\mathrm{AdS}_3/\mathbb{Z}$. As $\rho \to \infty$ for each slice of $\mathrm{AdS}_3/\mathbb{Z}$, the conformal boundary is characterized by a Lorentzian torus denoted as $\mathcal{C}\mathcal{T}^2 = S^1 \times S^1$, often called the celestial torus. In the coordinates $\psi$ and $\phi$ on $\mathcal{C}\mathcal{T}^2$, the boundary metric is as follows
\begin{equation}
\mathrm{d} s_{\mathcal{I}}^2=-\mathrm{d} \psi^2+\mathrm{d} \phi^2.
\end{equation} 
Null coordinates are given by
\begin{equation}
    \mathfrak{Z} = \frac{\psi + \phi}{2}~,~\bar{\mathfrak{Z}} = \frac{\psi - \phi}{2}.
\end{equation}
%The Lorentz group of Klein space $\mathrm{SL}(2, \mathbb{R})_{\text{left}}\cross  \mathrm{SL}(2,\mathbb{R})_{\text{right}}$ acts via independent M\"obius transformations on $z=\tan \mathfrak{Z}$ and $\bar{z}=\tan \bar{\mathfrak{Z}}$, forming the conformal group of $\mathcal{CT}^2$.
%where the conformal metric becomes $\mathrm{d}s^2 = -4\,\mathrm{d}\mathfrak{Z}\,\mathrm{d}\bar{\mathfrak{Z}}$. 
The Lorentz group of Klein space, denoted as $\mathrm{SL}(2, \mathbb{R})_{\text{left}} \times \mathrm{SL}(2,\mathbb{R})_{\text{right}}$, acts through independent M\"obius transformations on $z=\tan \mathfrak{Z}$ and $\bar{z}=\tan \bar{\mathfrak{Z}}$, thereby constituting the conformal group of $\mathcal{CT}^2$. Under this action, the conformal metric takes the form 
\begin{equation}
\mathrm{d}s_{\mathcal{I}}^2 = -4\,\mathrm{d}\mathfrak{Z}\,\mathrm{d}\bar{\mathfrak{Z}}.
\end{equation}
We also have
\begin{equation}
    (\mathfrak{Z},\bar{\mathfrak{Z}}) \sim (\mathfrak{Z} + \pi(m + n), \bar{\mathfrak{Z}} + \pi(m - n)), \qquad m, n \in \mathbb{Z}.
\end{equation}
\subsection{Celestial amplitudes in Klein space}
\label{ctor} 
In this subsection \ref{ctor}, we discuss celestial amplitudes in Klein space. For the discussion on the celestial amplitudes in Klein space, we closely follow \cite{Melton:2023bjw}. To analyze celestial amplitudes in Klein space, we employ the coordinates $\mathfrak{Z}$ and $\bar{\mathfrak{Z}}$, and 
\begin{equation}
(\mathfrak{Z},\bar{\mathfrak{Z}}) \sim (\mathfrak{Z}+2\pi,\bar{\mathfrak{Z}}),
\end{equation}
and 
\begin{equation}
(\mathfrak{Z},\bar{\mathfrak{Z}}) \sim (\mathfrak{Z}+\pi,\bar{\mathfrak{Z}}+\pi),
\end{equation}
covering the torus precisely once across the fundamental domain $\mathfrak{Z} \in [0,2\pi)$ and $\bar{\mathfrak{Z}} \in [0,\pi)$.

Massless momenta $q_i^\mu$ in $(2,2)$ signature can be parametrized by frequencies $\omega_i > 0$ and points $(\mathfrak{Z}_i,\bar{\mathfrak{Z}}_i) \in \mathcal{CT}^2$ as follows
\begin{align}
    q_i^\mu &= \omega_i \hat{q}_i^\mu, \\
    \hat{q}_i^\mu &= (\cos\psi_i, \sin\psi_i, \cos \phi_i, \sin \phi_i), \\
    \psi_i &:= \mathfrak{Z}_i + \bar{\mathfrak{Z}}_i, \quad \phi_i := \mathfrak{Z}_i - \bar{\mathfrak{Z}}_i.
\end{align}
We introduce the following shorthand notation for the separation of the torus
\begin{align}
    s_{ij} &:= \sin\mathfrak{Z}_{ij}, \\
    \bar{s}_{ij} &:= \sin\bar{\mathfrak{Z}}_{ij}.
\end{align}

If $A(1^{J_1}2^{J_2}\cdots n^{J_n})$ denotes the scattering amplitude of momentum eigenstates identified by null momenta $p_i$ and helicities $J_i$, the celestial amplitude is given by 
\begin{equation}
    \mathcal{A}(1^{J_1}2^{J_2}\cdots n^{J_n}) = \prod_{i=1}^n \int_0^\infty d\omega_i \, \omega_i^{\Delta_i-1} \, A(1^{J_1}2^{J_2}\cdots n^{J_n}).
\end{equation}
This undergoes transformation of a CFT correlator involving operators with conformal weights 
%This transforms as a CFT correlator of operators with conformal weights
\begin{equation}
    (h_i, \bar{h}_i) = \left(\frac{\Delta_i + J_i}{2}, \frac{\Delta_i - J_i}{2}\right),
\end{equation}
in a $2d$ CFT residing on $\mathcal{CT}^2$.
%Massless particles with spin $s$ are associated with appropriate polarization spinors, vectors, or tensors. 
\subsection{Celestial leaf amplitudes}
For the discussion on the celestial leaf amplitudes, we closely follow \cite{Melton:2023bjw}.
We are interested in exploring a refinement of celestial amplitudes known as leaf amplitudes, which are inspired by the hyperbolic foliation of Klein space.

Klein space $\mathbb{R}^{2,2}$ can be split into a timelike wedge where $X^2<0$ and a spacelike wedge where $X^2>0$. In each of these wedges, we can represent $X^\mu=\tau\hat{x}^\mu$, with $\tau>0$ and $\hat{x}^2=-1$ or $+1$. Constant $\tau$ slices in both scenarios correspond to copies of Lorentzian $\text{AdS}_3/\mathbb{Z}$. Together, these slices foliate Klein space.
In this context, we focus on celestial leaf amplitudes contributing to tree-level MHV gluon amplitudes, defined by the color-stripped Parke-Taylor formula
\begin{equation}
    A(1^-2^-3^+\cdots n^+) = \frac{\langle 12\rangle^3}{\langle 23\rangle\langle 34\rangle\cdots\langle n1\rangle}~\delta^4\bigg(\sum_{i=1}^n p_i\bigg) = \frac{\langle 12\rangle^3}{\langle 23\rangle\langle 34\rangle\cdots\langle n1\rangle}\int_{\mathbb{R}^{2,2}}\frac{\mathrm{d}^4X}{(2\pi)^4}~\mathrm{e}^{i\sum_jp_j\cdot X},
\end{equation}
where it is expressed in terms of the spinor-helicity brackets
\begin{equation}
    \langle ij\rangle := \sqrt{\omega_i\omega_j}\,s_{ij}.
\end{equation}
By substituting the Fourier representation of the momentum-conserving delta function and dividing it into integrals over the timelike and spacelike wedges, we integrate out $\tau$ (the magnitude of $X^\mu$) and perform Mellin transforms to obtain
\begin{equation}
    \mathcal{A}(1^-2^-3^+\cdots n^+) = \frac{\delta(\beta)}{8\pi^3}\,(\mathfrak{L}(\mathfrak{Z}_i,\bar{\mathfrak{Z}}_i) + \mathfrak{L}(\mathfrak{Z}_i,-\bar{\mathfrak{Z}}_i)),
\end{equation}
where $\beta=\sum_i(\Delta_i-1)$. This decomposes the celestial amplitude into building blocks defined along the slices (or leaves) of the hyperbolic foliation
\begin{equation}\label{leaf}
\mathfrak{L}(\mathfrak{Z}_i,\bar{\mathfrak{Z}}_i) = \frac{s_{12}^3}{s_{23}s_{34}\cdots s_{n1}}\int_{\hat x^2=-1}\mathrm{D}^3\hat x\;\prod_{i=1}^n\frac{\Gamma(2\bar{h}_i)}{(-i\hat p_i\cdot\hat x+\epsilon)^{2\bar{h}_i}},
\end{equation}
where $\epsilon > 0$ is a small regulator for the convergence of Mellin integrals, $\mathrm{D}^3\hat x$ is $\text{AdS}_3/\mathbb{Z}$ measure, and $\mathfrak{L}(\mathfrak{Z}_i,\bar{\mathfrak{Z}}_i) := \mathfrak{L}(1^-2^-3^+\cdots n^+)$ is referred to as the celestial leaf amplitude.

The Celestial leaf amplitudes can be expressed as correlators of spin-$1$ conformal primaries $\mathcal{O}^-_{\Delta_i}(\mathfrak{Z}_i,\bar{\mathfrak{Z}}_i)$
\begin{equation}
    \mathfrak{L}(1^-2^-3^+\cdots n^+) = \langle \mathcal{O}^-_{\Delta_1}(\mathfrak{Z}_1,\bar{\mathfrak{Z}}_1)\, \mathcal{O}^-_{\Delta_2}(\mathfrak{Z}_2,\bar{\mathfrak{Z}}_2) \cdots \mathcal{O}^+_{\Delta_n}(\mathfrak{Z}_n,\bar{\mathfrak{Z}}_n)\rangle_{\mathfrak{L}}.
\end{equation}
%\section{Spectral representation in $\mathrm{AdS}_3/\mathbb{Z}$ space}
%The Klein space can be foliated by $\mathrm{AdS}_3/\mathbb{Z}$, similar to how de Sitter (dS) spacetime can be sliced. Just as with dS spacetime, there is both a continuous and a discrete part to the identity resolution on $\mathrm{AdS}_3/\mathbb{Z}$.
%The unit metric on $\mathrm{AdS}_{3}/\mathbb{Z}$ is given by
%\begin{equation} 
%	ds_{3}^{2}=-\cosh^{2}\rho \, d\psi^{2}+\sinh^{2}\rho \, d\phi^{2}+d\rho^{2}.
%\end{equation}
%The $\mathbb{Z}$ acts as the time-like quotient $\psi \sim \psi+2\pi$.
%The resolution of identity on $\mathrm{AdS}_{3}/\mathbb{Z}$ contains both continuous representations from the $\phi$ and $\rho$ coordinates, and discrete representations from the $\psi$ coordinate. This is analogous to the two types of modes in dS spacetime, where there are continuous modes from the non-compact spatial coordinates, and discrete modes from the periodic time coordinate.
 
%\subsection{Harmonic analysis on $\mathrm{AdS}_3/\mathbb{Z}$}
%The global coordinates of $\mathrm{AdS}_3/\mathbb{Z}$ spacetime, 

%\section{The Resolution of Identity}

\section{Plancherel measure and resolution of identity in $\mathrm{AdS}_3/\mathbb{Z}$}
\label{dc}
In this section \ref{dc}, we study the Plancherel measure and resolution of identity in $\mathrm{AdS}_3/\mathbb{Z}$ to compute the spectral representation in Klein space. 
First we note that 
\begin{equation}
\operatorname{AdS}_{d+1} / \mathbb{Z}=\mathrm{H}_{d,2}.
\end{equation}
From a group theory perspective, we can represent hyperbolic space $\mathrm{H}_{n}$ as the quotient space $\mathrm{SO}(n,1)/\mathrm{SO}(n)$. Real hyperbolic space $\mathrm{H}_{p,q}$ can be expressed as
\begin{equation}
\mathrm{H}_{p, q}=\mathrm{SO}(p, q) / \mathrm{SO}(p, q-1),
\end{equation}
for $p \geqslant 1$ and can be embedded as the hypersurface
\begin{equation}
X \cdot X = -X_1^2 - \cdots - X_p^2 + X_{p+1}^2 + \cdots + X_{p+q}^2 = 1.
\end{equation}
%in the ambient space $X \in \mathbb{R}^{p, q}$.
Our hyperbolic space of interest is $\mathrm{AdS}_3/\mathbb{Z}$, i.e. $d=2$ , $p=2$ and $q=2$, all even
\begin{equation}
\mathrm{H}_{2, 2} = \frac{\mathrm{SO}(2, 2)}{\mathrm{SO}(2,1)} = G/H.
\end{equation}
Typically, the homogeneous space $\mathrm{SO}(p,q)/\mathrm{SO}(p,q-1)$ is represented by the hyperboloid $\mathrm{H}_{p,q}=\{(x,t)\in {\mathbb{R}^{p, q}}:\abs{x}^2-\abs{t}^2=1\}$. We note that for  Euclidean AdS space, and dS spacetime 
\begin{equation}
\begin{split}
&\mathrm{EAdS}_{d+1}=\mathrm{H}_{d+1,1}=\mathrm{SO}(d+1,1)/\mathrm{SO}(d+1)\\
&\mathrm{dS}_{d+1}=\mathrm{H}_{1,d+1}=\mathrm{SO}(1,1+d)/\mathrm{SO}(1, d).
\end{split}
\end{equation}
Now, we introduce hyperbolic coordinates in the region where $\abs{x}^2-\abs{t}^2>0$ as follows
\begin{equation}
\abs{x}^2-\abs{t}^2=m^2,  \quad m>0 \quad \text{and} ~~~h:=(h_t,h_x)=\left(\frac{t}{m}, \frac{x}{m}\right)\in \mathrm{H}.
\end{equation}
We can now write the laplacian operator $\square=-\Delta_t+\Delta_x$ in coordinates $t$ and $x$ as
\begin{equation}
\square=\frac{\partial^2}{\partial m^2}+\frac{p+q-1}{m} \frac{\partial}{\partial m}-\frac{1}{m^2} \square_h,
\end{equation}
where $\square_h$ is a second order partial differential operator on $\mathrm{H}$ which commutes with the action of $SO(p, q)$.

%The operator $\square_h$ is self-adjoint and so admits a spectral decomposition. This spectral decomposition also gives us the decomposition of the representation of $SO(p,q)$ on $L^2(\text{H})$.

The operator \(\square_h\) is self-adjoint, allowing for a spectral decomposition. This decomposition also gives a split of the representation of \(O(p, q)\) on \(L^2(\mathrm{H})\) (aside from a trivial decomposition into even and odd functions). The key observation is that if \(f(h)\) is an eigenfunction of \(\square_h\) with eigenvalue \(\lambda\), such that \(\square_h f = \lambda f\), then \(m^\alpha f(h)\) is a homogeneous solution (of degree \(\alpha\)) to \(\square u = 0\) in the region \(|x| > |t|\), given that \(\lambda = \alpha(\alpha + p + q - 2)\). In other words
\begin{equation}
\alpha = -\frac{p+q-2}{2} \pm \sqrt{\left[\frac{p+q-2}{2}\right]^2 + \lambda}.
\end{equation}
We have
\begin{equation}
u(m, h) = m^{-\frac{p+q-2}{2} \pm \sqrt{B}} f(h),
\end{equation}
where \(B = \left(\frac{p+q-2}{2}\right)^2 + \square_h\), is a solution to \(\square u = 0\) in \(|x| \geq |t|\) for any \(f \in L^2(\mathrm{H}) \). The challenge is that \(B\) is not a positive operator, making \(\sqrt{B}\) not well-defined. To address this, we decompose \(B\) (and therefore \(L^2(\mathrm{H})\)) into its positive and negative parts. Denoting by \(P(a, b)\) the spectral projection of \(B\) onto the interval \((a, b]\), and \(\mathscr{H}(a, b) = P(a, b) (L^2(\mathrm{H}) \), the restriction of \(B\) to \(\mathscr{H}(-\infty, 0)\) and \(\mathscr{H}(0, \infty)\) is negative and positive, respectively. %We first examine the spectrum of \(B\) in \((0, \infty)\).

The spectrum of the operator $\square_{h}$ consists of a discrete part (except when $q=1$), \footnote{In the context of $\mathrm{EAdS}_{d+1}$ space (Euclidean AdS space), only a continuous spectrum is present. Here, for $\mathrm{EAdS}_{d+1}, ~q=1$, the quotient part $\mathrm{SO}(p,q-1)$ reduces to $\mathrm{SO}(p),~p=d+1$, i.e. $\mathrm{SO}(d+1,1)/\mathrm{SO}(d+1)$.} alongside continuous part $\mathscr{H}_{-}:=\mathscr{H}{(-\infty,0)}$. When viewed as a representation of $\mathrm{SO}(p,q)$, $\mathscr{H}_{-}\bigoplus \mathscr{H}_{-}$ is unitarily equivalent to the regular representation on $L^2(\mathrm{C})$, where $\mathrm{C}$ is the cone 
\begin{equation}
\{(x,t)\in {\mathbb{R}^{p, q}}:\abs{x}^2-\abs{t}^2=0\},
\end{equation}
both $\mathrm{H}$ and $\mathrm{C}$ have unique invariant measures under the action of $\mathrm{SO}(p,q)$. \\

The invariant measure on hyperboloid $\mathrm{H}$ tends to invariant measure on $\mathrm{C}$ when we take the limit $m\to 0$, implying that we obtain a unitary operator from  $\mathscr{H}_{-}\bigoplus\mathscr{H}_{-}$ 
to $L^2(\mathrm{C})$.

\paragraph{Intertwining Operator.}

There are two equivalent unitary representation of $\mathrm{SO}(p,q)$, one is on $L^2(\mathrm{C})$ and another on $L^2(\mathrm{H}) $. They intertwine the $\mathrm{SO}(p,q)$ group action. The  morphisms between two different representation frequently called as intertwining operator. We now discuss intertwining operator for $\mathrm{SO}(p,q)$. Let us introduce coordinates $x^{\prime}$, $t^{\prime}$, and $s$ in $\mathrm{C}$ as follows 
$$x^{\prime}=x /|x| \in S^{p-1},~t^{\prime}=t /|t|=S^{q-1},$$ and $$s=|x|=|t|,~\text{where}~0<s<\infty.$$ The invariant measure in this space is $s^{p+q-3} \mathrm{d} x^{\prime} \mathrm{d} t^{\prime} \mathrm{d} s$. Any arbitrary function within $L^2(\mathrm{C})$ can be expressed as
\begin{equation}
F\left(x^{\prime}, t^{\prime}, s\right)=\frac{1}{2 \pi} \int_{-\infty}^{\infty} \varphi\left(x^{\prime}, t^{\prime}, \rho\right) s^{-\frac{p+q-2}{2}+i \rho} \mathrm{d} \rho,
\end{equation}
and
\begin{equation}
\|F\|^2=\frac{1}{2 \pi} \int_{-\infty}^{\infty} \iint\left|\varphi\left(x^{\prime}, t^{\prime}, \rho\right)\right|^2 \mathrm{d} x^{\prime} \mathrm{d} t^{\prime} \mathrm{d}\rho .
\end{equation}
%\textbf{Theorem.} 
Now, for each $\rho \neq 0$ there exist positive constants $a(\rho)$ and $b(\rho)$ such that
\begin{equation}
A(\rho) \psi\left(x^{\prime}, t^{\prime}\right)=a(\rho) \int_{S^{p-1}} \int_{S^{q-1}}\left|x^{\prime} \cdot \xi^{\prime}-t^{\prime} \cdot \tau^{\prime}\right|^{-\frac{p+q-2}{2}+i \rho} \psi\left(\xi^{\prime}, \tau^{\prime}\right) \mathrm{d} \tau^{\prime} \mathrm{d} \xi^{\prime},
\end{equation}
is a unitary operator on even functions in $L^2\left(S^{p-1} \times S^{q-1}\right)$ and
\begin{equation}
\begin{aligned}
B(\rho) \psi\left(x^{\prime}, t^{\prime}\right)= & b(\rho) \int_{S^{p-1}} \int_{S^{q-1}}\left|x^{\prime} \cdot \xi^{\prime}-t^{\prime} \cdot \tau^{\prime}\right|^{-\frac{p+q-2}{2}+i \rho} \\
& \times \operatorname{sgn}\left(x^{\prime} \cdot \xi^{\prime}-t^{\prime} \cdot \tau^{\prime}\right) \psi\left(\xi^{\prime}, \tau^{\prime}\right) \mathrm{d} \tau^{\prime} \mathrm{d} \xi^{\prime},
\end{aligned}
\end{equation}
is a unitary operator on odd functions in $L^2\left(S^{p-1} \times S^{q-1}\right)$.
Therefore, we can write
\begin{equation}
\frac{1}{2}\left(\varphi\left(x^{\prime}, t^{\prime}, \rho\right)+\varphi\left(-x^{\prime},-t^{\prime}, \rho\right)\right)=a(\rho) A(\rho) \psi_1,
\end{equation}
and
\begin{equation}
\frac{1}{2}\left(\varphi\left(x^{\prime}, t^{\prime}, \rho\right)-\varphi\left(-x^{\prime},-t^{\prime}, \rho\right)\right)=b(\rho) B(\rho) \psi_2.
\end{equation}
We can write any arbitrary function $f \in \mathscr{H}_{-}$ as 
\begin{equation}
\begin{aligned}
f(x, t)= & \frac{1}{2 \pi} \int_0^{\infty} \int_{S^{p-1}} \int_{S^{q-1}}\left|x \cdot \xi^{\prime}-t \cdot \tau^{\prime}\right|^{-\frac{p+q-2}{2}+i \rho} \\
& \times \psi_1\left(\xi^{\prime}, \tau^{\prime}, \rho\right) \mathrm{d} \xi^{\prime} \mathrm{d} \tau^{\prime} a(\rho)^2 \mathrm{d} \rho \\
& +\frac{1}{2 \pi} \int_0^{\infty} \int_{S^{p-1}} \int_{S^{q-1}}\left|x \cdot \xi^{\prime}-t \cdot \tau^{\prime}\right|^{-\frac{q+p-2}{2}+i \rho} \\
& \times \operatorname{sgn}\left(x \cdot \xi^{\prime}-t \cdot \tau^{\prime}\right) \psi_2\left(\xi^{\prime}, \tau^{\prime}, \rho\right) \mathrm{d} \xi^{\prime} \mathrm{d} \tau^{\prime} b(\rho)^2 \mathrm{d} \rho,
\end{aligned}
\end{equation}
with
\begin{equation}
\begin{aligned}
\|f\|^2= & \frac{1}{2 \pi} \int_0^{\infty} \int_{S^{p-1}} \int_{S^{q-1}}\left|\psi_1\left(\xi^{\prime}, \tau^{\prime}, \rho\right)\right|^2 \mathrm{d} \xi^{\prime} \mathrm{d} \tau^{\prime} a(\rho)^2 \mathrm{d} \rho \\
& +\frac{1}{2 \pi} \int_0^{\infty} \int_{S^{p-1}} \int_{S^{q-1}}\left|\psi_2\left(\xi^{\prime}, \tau^{\prime}, \rho\right)\right|^2 \mathrm{d} \xi^{\prime} \mathrm{d} \tau^{\prime} b(\rho)^2 \mathrm{d} \rho .
\end{aligned}
\end{equation}
We have 
\begin{equation}
\psi_1\left(\xi^{\prime}, \tau^{\prime}, \rho\right)=\int_{\mathrm{H}}\left|x \cdot \xi^{\prime}-t \cdot \tau^{\prime}\right|^{-\frac{p+q-2}{2}-i \rho}\left(x \cdot \xi^{\prime}-t \cdot \tau^{\prime}\right) f(x, t) \mathrm{d} h,
\end{equation}
and
\begin{equation}
\psi_2\left(\xi^{\prime}, \tau^{\prime}, \rho\right)=\int_{\mathrm{H}}\left|x \cdot \xi^{\prime}-t \cdot \tau^{\prime}\right|^{-\frac{p+q-2}{2}-i \rho} \operatorname{sgn}\left(x \cdot \xi^{\prime}-t \cdot \tau^{\prime}\right) f(x, t) \mathrm{d} h.
\end{equation}
%if the integrals are interpreted in a suitable sense. 
%The resulting expansion decomposes $\mathscr{H}(-\infty, 0)$ into a direct integral of irreducible unitary representations of $\mathrm{\mathrm{SO}}(p, q)$. 
The measures $a(\rho)^2 \mathrm{d} \rho$ and $b(\rho)^2 d \rho$ represent the Plancherel measures. The determination of $a(\rho)$ and $b(\rho)$ involves evaluating $A(\rho)$ and $B(\rho)$ for a particular function.

In the case of $\mathrm{SO}(2,2)$ acting on the hyperboloid, we can give an interesting interpretation. The quadratic expression $-t_1^2-t_2^2+x_1^2+x_2^2$ can be readily recognized as equivalent to $a d-b c$, the determinant of a $2 \times 2$ matrix. Consequently, the hyperboloid can be associated with $\mathrm{SL}(2,\mathbb{R})$, and the group's action is just $\mathrm{SL}(2,\mathbb{R})$ acting on itself both from the left and the right. Ultimately, this yields the Plancherel formula for $\mathrm{SL}(2,\mathbb{R})$.

Now, we compute $a(\rho), \text{and} ~b(\rho)$. We have
\begin{equation}
A(\rho) \psi(x', t') = a(\rho) \int_{S^{q-1}} \int_{S^{p-1}} \left| x' \cdot \xi' - t' \cdot \tau' \right|^{-\frac{p+q-2}{2} + i\rho} \psi(\xi', \tau') \,\mathrm{d} \tau^{\prime}\, \mathrm{d}\xi',
\end{equation}
and
\begin{equation}
\begin{split}
B(\rho) \psi(x', t') &= b(\rho) \int_{S^{q-1}} \int_{S^{p-1}} \left| x' \cdot \xi' - t' \cdot \tau' \right|^{-\frac{p+q-2}{2} + i\rho}\\
&\qquad\qquad\qquad\qquad\quad \times \text{sgn}(x' \cdot \xi' - t' \cdot \tau') \psi(\xi', \tau') \, \mathrm{d} \tau^{\prime}\,\mathrm{d}\xi'.
\end{split}
\end{equation}
We choose $$\psi(x', t') = C_k^{\alpha}(\xi' \cdot \eta') ~C_j^{\beta}(\tau' \cdot \sigma')~ \text{ where } \alpha =\frac{(p-2)}{2} \text{ and} ~\beta=\frac{(q-2)}{2}.$$  Here $C_k^{\alpha}$,~$C_j^{\beta}$ are the Gegenbauer polynomials
\begin{equation}
\begin{split}
C_k^{\alpha}(t) &= d_{k, \alpha}(1 - t^2)^{\alpha - \frac{1}{2}}\left(\frac{\mathrm{d}}{\mathrm{d}t}\right)^k (1 - t^2)^{k + \alpha - \frac{1}{2}}, \\
d_{k, \alpha} &= \frac{\Gamma(\alpha + \frac{1}{2}) \Gamma(2\alpha + k)}{(-2)^k \Gamma(k + 1) \Gamma(\alpha + k + \frac{1}{2}) \Gamma(2\alpha)}.
\end{split}
\end{equation}
We get 
\begin{equation}
\begin{split}
a(\rho)^{-1} A(\rho)\psi(x', t') &= \mathfrak{g}(\rho, k, j, p, q)\psi(x', t') \\
\text{if } k + j \text{ is even},
\end{split}
\end{equation}
\text{and}
\begin{equation}
\begin{split}
b(\rho)^{-1} B(\rho)\psi(x', t') &= \mathfrak{g}(\rho, k, j, p, q)\psi(x', t') \\
\text{if } k + j \text{ is odd}.
\end{split}
\end{equation}

\paragraph{Case $p=q= \text{even}$.}
Now, we compute some necessary quantities as a preparation to compute the Plancherel measure.
We choose to integrate at $x^{\prime}= \eta^{\prime}$ and $t^{\prime} = \sigma^{\prime}$
\begin{equation}
\begin{aligned}
\mathfrak{g}(\rho, k, j, p, q)= & e_{k, \alpha} e_{j, \beta} \int_{-1}^1 \int_{-1}^1\left|t_1-t_2\right|^{-\frac{p+q-2}{2}+i \rho} \\
& \times\left[\operatorname{sgn}\left(t_1-t_2\right)\right]^{k+j}\left(\frac{\mathrm{d}}{\mathrm{d} t_1}\right)^k\left(1-t_1^2\right)^{k+\lambda-\frac{1}{2}} \\
& \times \left(\frac{d}{d t_2}\right)^j\left(1-t_2^2\right)^{j+\beta-\frac{1}{2}} \mathrm{d}t_1 \mathrm{d} t_2 ,
\end{aligned}
\end{equation}
where 
\begin{equation}
\begin{split}
&e_{k,\alpha}=\frac{\Omega_{p-1}~d_{k,\alpha}}{C^\alpha_k(1)}, e_{j,\beta}=\frac{\Omega_{q-1}~d_{k,\beta}}{C^\beta_j(1)}, \\
&t_1=\cos{\theta_1}, t_2=\cos{\theta_2}, \\
&\Omega_{p-1}=\frac{2\pi^{\frac{p-1}{2}}}{\Gamma(\frac{p-1}{2})} 
,~\text{and} ~\Omega_{q-1}=\frac{2\pi^{\frac{q-1}{2}}}{\Gamma(\frac{q-1}{2})}.
\end{split}
\end{equation}
If $m \leqslant \beta$ we integrate by parts in the $t_2$-variables $m$ times to obtain
\begin{equation}
\begin{aligned}
& e_{k, \alpha} e_{j, \beta} \int_{-1}^1 \int_{-1}^1 \frac{\Gamma\left(1-\frac{p+q-2}{2}+i \rho\right)}{\Gamma\left(1+m-\frac{p+q-2}{2}+i \rho\right)}\left|t_1-t_2\right|^{m-\frac{p+q-2}{2}+i \rho} \\
& \times\left[\operatorname{sgn}\left(t_1-t_2\right)\right]^{k+j+m}\left(\frac{\mathrm{d}}{\mathrm{d} t_1}\right)^k\left(1-t_1^2\right)^{k+\alpha-\frac{1}{2}} \\
& \times \left(\frac{\mathrm{d}}{\mathrm{d} t_2}\right)^{m+j}\left(1-t_2^2\right)^{j+\beta-\frac{1}{2}} \mathrm{d} t_1 \mathrm{d} t_2 ,
\end{aligned}
\end{equation}
and therefore
\begin{equation}
\begin{aligned}
\mathfrak{g}(\rho, k, j, p, q)= & \frac{e_{j, \beta}}{e_{m+j, \beta-m}} \frac{\Gamma\left(1-\frac{p+q-2}{2}+i \rho\right)}{\Gamma\left(1+m-\frac{p+q-2}{2}+i \rho\right)} ~ \mathfrak{g}(\rho, k, j+m, p, q-2 m) .
\end{aligned}
\end{equation}
Now we first choose $m=\beta$ then interchange the role of $p$ and $q$ and apply the above relation with $m=\alpha$, and finally we get 
\begin{equation}
\begin{aligned}
\mathfrak{g}(\rho, k, j, p, q)= & \frac{e_{j, \beta}}{e_{\beta+j, 0}} \cdot \frac{e_{k, \alpha}}{e_{k+\alpha, 0}} \frac{\Gamma\left(1-\frac{p+q-2}{2}+i \rho\right)}{\Gamma(1-p / 2+i \rho)} \frac{\Gamma\left(1-\frac{p}{2}+i \rho\right)}{\Gamma\left(i\rho\right)} ~\mathfrak{g}(\rho, k+\alpha, j+ \beta, 2,2) .
\end{aligned}
\end{equation}
We have $\left|e_{j, \beta} / e_{\beta+j, 0}\right|=(2 \pi)^{\beta}$ and
\begin{equation}
\left|\frac{\Gamma\left(1-\frac{p+q-2}{2}+i \rho\right)}{ \Gamma(i \rho)}\right|=\left|\frac{\Gamma(1+i \rho)}{\Gamma\left(\frac{p+q-2}{2}+i \rho\right)}\right|.
\end{equation}
We find
\begin{equation}
\begin{aligned}
& a(\rho)=\frac{1}{2}(2 \pi)^{-\left(\frac{p+q-2}{2}\right)}\left|\frac{\Gamma\left(\frac{p+q-2}{2}+i \rho\right)}{\Gamma(i \rho)}\right||\tanh \pi \rho / 2|, \\
& b(\rho)=\frac{1}{2}(2 \pi)^{-\left(\frac{p+q-2}{2}\right)}\left|\frac{\Gamma\left(\frac{p+q-2}{2}+i \rho\right)}{\Gamma(i \rho)}\right||\operatorname{coth} \pi \rho / 2| .
\end{aligned}
\end{equation}

\subsection{Computation of Plancherel measure in $\mathrm{AdS}_3/\mathbb{Z}$}
\label{planmeas}
In this section \ref{planmeas}, we compute the Plancherel measure, for the case of $p$ and $q$ both even with $p+q=d+2$.
We note that $\mathrm{AdS}_3/\mathbb{Z}$ comes with $p=2$ and $q=2$.
We will first discuss the case of $p$ and $q$ both even with $p+q=d+2$. 

$\mathrm{AdS}_{d+1}/\mathbb{Z}$ can be written as
$\mathrm{SO}(d,2)/\mathrm{SO}(d,1)$, i.e. $p=d$ and $q=2$ and $p+q=d+2$. The constants $a(\rho)$ and $b(\rho)$ can be computed as
\begin{equation}
\label{arho}
a(\rho)=\frac{1}{2}(2 \pi)^{-\frac{d}{2}} \frac{\left|\Gamma\left(\frac{d}{2}+i \rho\right)\right|}{|\Gamma(i \rho)|} {\left|\tanh{\frac{\pi \rho}{2}}\right|}
\end{equation}
and 
\begin{equation}
\label{brho}
b(\rho)=\frac{1}{2}(2 \pi)^{-\frac{d}{2}} \frac{\left|\Gamma\left(\frac{d}{2}+i \rho\right)\right|}{|\Gamma(i \rho)|} {\left|\coth{\frac{\pi \rho}{2}}\right|}
\end{equation}
Following \cite{Strichartz1989HarmonicAO, Rossmann}, any function \( f \) in \( L^2(\mathrm{H}_{p, q}) \) can be expressed as
\begin{equation}
\begin{split}
f &= \int_0^{\infty} \mathscr{P}_\lambda f \, d \lambda \quad + \sum_{j>-\rho} \mathscr{Q}_{i(j+\rho)} f,
\end{split}
\end{equation}
where \( j \) is an integer parameter. The operators \( \mathscr{P}_i \) and \( \mathscr{Q}_{i(j+\rho)} \) satisfy the following equations
\begin{equation}\label{discreteconttt}
\begin{split}
\square \mathscr{P}_i f &= \left(-\lambda^2 - \rho^2\right) \mathscr{P}_i f, \\
\square \mathscr{Q}_{i(j+\rho)} f &= \left((j+\rho)^2 - \rho^2\right) \mathscr{Q}_{i(j+\rho)} f.
\end{split}
\end{equation}
From \eqref{discreteconttt}, we can set $\frac{d}{2}=\lambda$ inside the $\Gamma$ function and the eqs.\eqref{arho}, and \eqref{brho} become
\begin{equation}
a(\rho)=\frac{1}{2}(2 \pi)^{-\frac{d}{2}} \frac{\left|\Gamma\left(\lambda+i \rho\right)\right|}{|\Gamma(i \rho)|}{\left|\tanh{\frac{\pi \rho}{2}}\right|}.
\end{equation}
and 
\begin{equation}
b(\rho)=\frac{1}{2}(2 \pi)^{-\frac{d}{2}} \frac{\left|\Gamma\left(\lambda+i \rho\right)\right|}{|\Gamma(i \rho)|}{\left|\coth{\frac{\pi \rho}{2}}\right|}
\end{equation}
Again looking at eq.\eqref{strum} (for details we refer to appendix \ref{sleqnn}) we can write $(\lambda+i \rho)$ as $\Delta$ and $(\lambda-i \rho)$ as $\tilde{\Delta}$ so that 
\begin{equation}\label{deltatilde}
    \lambda^2+\rho^2=\Delta (\Delta-d)=-\Delta \tilde{\Delta},
\end{equation}
Plancherel measure can be calculated as
\begin{equation}
    \frac{1}{\mu(\Delta)}=a(\rho)^2=\frac{1}{2^2}~ 2^{-d} ~\pi^{-d} \frac{\Gamma\left(\lambda+i \rho\right)}{\Gamma(i \rho)} \frac{\Gamma\left(\lambda- i\rho\right)}{\Gamma(-i \rho)} \tanh{\frac{\pi \rho}{2}}\tanh{\frac{\pi \rho^*}{2}}
\end{equation}
Now to write it in terms of $\Delta$ and $\tilde{\Delta}$ observe $\frac{d}{2}+i\rho=\Delta$. Using this we can write the Plancherel measure for even $p$ and even $q$ case with $p+q=d+2$ as
\begin{equation}
\small
    \frac{1}{\mu(\Delta)}=a(\rho)^2=\frac{1}{2^2}~ 2^{-d} ~\pi^{-d} \frac{\Gamma(\Delta)}{\Gamma\left(\Delta-\frac{d}{2}\right)} \frac{\Gamma(\tilde{\Delta)}}{\Gamma\left(\Delta-\frac{d}{2}\right)} \tan{\frac{\pi}{2}\left(\Delta-\frac{d}{2}\right)}\tan{\frac{\pi}{2}\left(\tilde{\Delta}-\frac{d}{2}\right)}.
\end{equation}
Similarly one can write it with $\cot$ as well for the odd function
\begin{equation}
\small
    \frac{1}{\mu(\Delta)}=b(\rho)^2=\frac{1}{2^2}~ 2^{-d} ~\pi^{-d} \frac{\Gamma(\Delta)}{\Gamma\left(\Delta-\frac{d}{2}\right)} \frac{\Gamma(\tilde{\Delta)}}{\Gamma\left(\Delta-\frac{d}{2}\right)} \cot{\frac{\pi}{2}\left(\Delta-\frac{d}{2}\right)}\cot{\frac{\pi}{2}\left(\tilde{\Delta}-\frac{d}{2}\right)}.
\end{equation}

Now coming back to our case of interest, the $\mathrm{AdS}_3/\mathbb{Z}$, we can calculate the Plancherel measure using \eqref{arhobrho} and with $\lambda=\frac{d}{2}=1$ value. 
The result is
\begin{equation}
\label{mudelta}
    \frac{1}{\mu(\Delta)}=a(\rho)^2
    %=\frac{1}{2^2}~ 2^{1-3} ~\pi^{1-3} \tan{\frac{\pi}{2}\left(\Delta-1\right)}\tan{\frac{\pi}{2}\left(\tilde{\Delta}-1\right)}
    =\frac{1}{16 \pi^2}  \tan{\frac{\pi}{2}\left(\Delta-1\right)}\tan{\frac{\pi}{2}\left(\tilde{\Delta}-1\right)},
\end{equation}
for even functions and 
\begin{equation}
    \frac{1}{\mu(\Delta)}=b(\rho)^2
    %=\frac{1}{2^2}~ 2^{1-3} ~\pi^{1-3} \cot{\frac{\pi}{2}\left(\Delta-1\right)}\cot{\frac{\pi}{2}\left(\tilde{\Delta}-1\right)}
    =\frac{1}{16 \pi^2} \cot{\frac{\pi}{2}\left(\Delta-1\right)}\cot{\frac{\pi}{2}\left(\tilde{\Delta}-1\right)},
\end{equation}
for odd functions.
Here, the shadow weight is
\begin{equation}
\tilde{\Delta}=2-\Delta. \nonumber
\end{equation}
In the appendix \ref{alder}, we give an alternative derivation of the Plancherel measure as a special case: case $p=q=2$.
%After simplification for even, and odd functions the Plancherel measure becomes
%\begin{equation}
%\begin{split}
%&\frac{1}{\mu(\Delta)}=a(\rho)^2=-\frac{\cot^2{\left(\frac{\pi \Delta}{2}\right)}}{16 \pi^2}\\
%&\frac{1}{\mu(\Delta)}=b(\rho)^2=-\frac{\tan^2{\left(\frac{\pi \Delta}{2}\right)}}{16 \pi^2}.
%\end{split}
%\end{equation}
\subsubsection{Relation to the Plancherel measure of the CLM prescription \cite{Chang:2023ttm}: Euclidean AdS space and dS spacetime}
At this stage, it might be helpful for our reader if we take a step back to connect with the Plancherel measure of the CLM prescription \cite{Chang:2023ttm}: Euclidean AdS space and dS spacetime.
Let us compute the plancheral measure of Euclidean AdS space and dS spacetime using our prescription. 
\subsection*{i. $\mathrm{EAdS}_{d+1}$ Plancherel measure.}
$\mathrm{EAdS}_{d+1}$ can be realized as $\mathrm{EAdS}_{d+1}=\mathrm{SO}(d+1,1)/\mathrm{SO}(d+1)$. The spectrum contains only continuous part. 
Our analysis belongs to the case $p=d+1$ and $q=1$. The constants $a(\rho)$ and $b(\rho)$ can be computed as
\begin{equation}
\label{abrho}
a(\rho)=b(\rho)=2(2 \pi)^{-\frac{d}{2}} \frac{\left|\Gamma\left(\frac{d}{2}+i \rho\right)\right|}{|\Gamma(i \rho)|}.
\end{equation}
For $\mathrm{H}_{p, q}$ the spectrum of $\square$ consists of a discrete part above $-\rho^2$ and a continuous part below $-\rho^2$, where $\rho=\frac{1}{2}(p+q-2)$. Following \cite{Strichartz1989HarmonicAO, Rossmann}, any function $f$ in $L^2\left(\mathrm{H}_{p, q}\right)$ can be represented as
\begin{equation}
\begin{split}
f &= \int_0^{\infty} \mathscr{P}_\lambda f \, d \lambda \quad + \sum_{j>-\rho} \mathscr{Q}_{i(j+\rho)} f,
\end{split}
\end{equation}
where $j$ is an integer parameter, and the operators $\mathscr{P}_i$ and $\mathscr{Q}_{i(j+\rho)}$ satisfy
\begin{equation}\label{discretecont1}
\begin{split}
\square \mathscr{P}_i f &= \left(-\lambda^2-\rho^2\right) \mathscr{P}_i f \\
\square \mathscr{Q}_{i(j+\rho)} f &= \left((j+\rho)^2-\rho^2\right) \mathscr{Q}_{i(j+\rho)} f.
\end{split}
\end{equation}
From eq.\eqref{discretecont1}, we can set $\lambda=\frac{d}{2}$ inside the $\Gamma$ function and the eq.\eqref{abrho} becomes
\begin{equation}
a(\rho)=b(\rho)=2(2 \pi)^{-\frac{d}{2}} \frac{\left|\Gamma\left(\lambda+i \rho\right)\right|}{|\Gamma(i \rho)|}.
\end{equation}
Now, looking at the Sturm-Liouville eq.\eqref{strum} (for details we refer to appendix \ref{sleqnn}) we can write $(\lambda+i \rho)$ as $\Delta$ and $(\lambda-i \rho)$ as $\tilde{\Delta}$ so that 
\begin{equation}\label{deltatilde}
    \lambda^2+\rho^2=\Delta (\Delta-d)=-\Delta \tilde{\Delta}.
\end{equation}
The Plancherel measure can be calculated as
\begin{equation}
    \frac{1}{\mu(\Delta)}=a(\rho)^2=b(\rho)^2=2^2~ 2^{-d} ~\pi^{-d} \frac{\Gamma\left(\lambda+i \rho\right)}{\Gamma(i \rho)} \frac{\Gamma\left(\lambda- i\rho\right)}{\Gamma(-i \rho)}.
\end{equation}
It should be noted here that 
\begin{equation}
    \overline{\Gamma(x)}=\Gamma(\bar{x}).
\end{equation}
Now  using eq.\eqref{deltatilde} we can write Plancherel measure in terms of $\Delta$ and $\tilde{\Delta}$ as following
\begin{equation}
    \frac{1}{\mu(\Delta)}= 2^2~ 2^{-d} ~\pi^{-d} \frac{\Gamma(\Delta)}{\Gamma\left(\Delta-\frac{d}{2}\right)} \frac{\Gamma(\tilde{\Delta})}{\Gamma\left(\tilde{\Delta}-\frac{d}{2}\right)}.  
\end{equation}

\subsection*{ii. $\mathrm{dS}_{d+1}$ Plancherel measure.}
Now, $\mathrm{dS}_{d+1}$ can be realized as $\mathrm{dS}_{d+1}=\mathrm{SO}(1,d+1)/\mathrm{SO}(1,d)$. The spectrum contains both continuous and discrete part. Our analysis belongs to the case $p=1$ and $q=d+1$. \\
The same procedure applies as in the $\mathrm{EAdS}_{d+1}$ case for computing $a(\rho)$ and $b(\rho)$; we simply replace $p$ with $q$, and the calculations remain unchanged. We have
\begin{equation}
a(\rho)=b(\rho)=2(2 \pi)^{-\frac{d}{2}} \frac{\left|\Gamma\left(\frac{d}{2}+i \rho\right)\right|}{|\Gamma(i \rho)|}.
\end{equation}
and the Plancherel measure is given by 
\begin{equation}
    \frac{1}{\mu(\Delta)}= 2^2~ 2^{-d} ~\pi^{-d} \frac{\Gamma(\Delta)}{\Gamma\left(\Delta-\frac{d}{2}\right)} \frac{\Gamma(\tilde{\Delta})}{\Gamma\left(\tilde{\Delta}-\frac{d}{2}\right)}.  
\end{equation}
The plancheral measure of Euclidean AdS space and dS spacetime using our prescription correctly matches with the CLM prescription \cite{Chang:2023ttm}. This is a nice consistency check of our prescription. 

\subsection{Resolution of identity in $\mathrm{AdS}_3/\mathbb{Z}$}
\label{ires}
Now, in this subsection we study the resolution of identity in $\mathrm{AdS}_3/\mathbb{Z}$ slices. 
The spectral analysis of laplacian $\square$ on $\mathrm{H}_{p, q}$ is essentially \footnote{The term ``essentially'' is used because there is a further subdivision of eigenfunctions into even and odd parts as in \cite{Strichartz1989HarmonicAO, Rossmann}.} the decomposition of the standard representation of $O(p, q)$ on $L^2\left(\mathrm{H}_{p, q}\right)$. The spectrum of $\square$ is composed of a discrete part above $-\rho^2$ and a continuous part below $-\rho^2$, where $\rho=\frac{1}{2}(p+q-2)$. Following \cite{Strichartz1989HarmonicAO, Rossmann}, any function $f \in L^2\left(\mathrm{H}_{p, q}\right)$ can be expressed as
\begin{equation}
\begin{split}
f &= \int_0^{\infty} \mathscr{P}_\lambda f \, d \lambda \quad + \sum_{j>-\rho} \mathscr{Q}_{i(j+\rho)} f,
\end{split}
\end{equation}
where $j$ is an integer parameter, and the operators $\mathscr{P}_i$ and $\mathscr{Q}_{i(j+\rho)}$ satisfy
\begin{equation}\label{discretecont}
\begin{split}
\square \mathscr{P}_i f &= \left(-\lambda^2-\rho^2\right) \mathscr{P}_i f \\
\square \mathscr{Q}_{i(j+\rho)} f &= \left((j+\rho)^2-\rho^2\right) \mathscr{Q}_{i(j+\rho)} f.
\end{split}
\end{equation}
The hyperbolic space we are focusing on is $\mathrm{AdS}_3/\mathbb{Z}$, where we have $p=2$, and $q=2$. 
Now, we choose function $f(P)$, and $f(X)$ 
\begin{equation}
F_{\Delta, \epsilon}^{-1} : f(P) \mapsto \int_{\mathrm{PC}_{2,2}} \mathrm{d} P |X \cdot P|^{-\Delta, \epsilon} f(P) \in L^2(\mathrm{AdS}_3/\mathbb{Z}).
\end{equation}
Here, we use the shorthand notation 
\begin{equation}
|X \cdot P|^{-\Delta, \epsilon}:=|X \cdot P|^{-\Delta}\text{sgn}^{\epsilon}(X \cdot P).
\end{equation}
Here, the lightcone $P \cdot P = 0$ for $P \neq 0$ and $P \in \mathbb{R}^{2,2}$ is denoted as $\text{LC}_{2,2}$. Consequently, the asymptotic boundary of $\text{H}_{2,2}$ is the projective lightcone $\text{PC}_{2,2}$, defined as 
\begin{equation}
\text{PC}_{2,2} := \text{LC}_{2,2} / \mathbb{R}^+ \cong S^{1} \times S^{1},
\end{equation}
under the equivalence relation $P \sim \lambda P$ for $\lambda > 0$.
We have 
\begin{equation}
F_{\Delta, \epsilon} : f(X) \mapsto \int_{\mathrm{AdS}_3/\mathbb{Z}} \mathrm{d}X |X \cdot P|^{-\tilde{\Delta}, \epsilon} f(X),
\end{equation}
\begin{equation}
\mathscr{P}_{\Delta, \epsilon} : f(X) \mapsto \int_{\mathrm{AdS}_3/\mathbb{Z}} \mathrm{d}X' \phi_{\Delta, \epsilon}(X, X') f(X').
\end{equation}
The inversion formula is
\begin{equation}
\begin{split}
f(X) &= \frac{1}{2\pi i} \sum_{\epsilon=0,1} \int_{\Gamma} \frac{\mathrm{d}\Delta}{\mu(\Delta)} \mathscr{P}_{\Delta, \epsilon}[f(X)] +  \sum_{\substack{(\Delta, \epsilon) \in D_{\text{}}}} \text{Res} \frac{1}{\mu(\Delta)} \mathscr{P}_{\Delta, \epsilon}[f(X)].
\end{split}
\end{equation}
The Dirac delta distribution's spectral representation on the $\mathrm{AdS}_3/\mathbb{Z}$ spacetime is given by
\begin{equation}
\label{riden}
\begin{split}
\delta(X, X') &= \frac{1}{2\pi i} \sum_{\epsilon=0,1} \int_{\Gamma} \frac{\mathrm{d}\Delta}{2\mu(\Delta)} \int_{\mathrm{PC}_{2,2}} \mathrm{d}P |2X \cdot P|^{-\tilde{\Delta}, \epsilon}|2X' \cdot P|^{-\Delta, \epsilon} \\
&\qquad\qquad +  \sum_{\substack{(\Delta, \epsilon) \in D_{\text{}}}} \text{Res} \int_{\mathrm{PC}_{2,2}} \mathrm{d}P \frac{1}{2\mu(\Delta)} |2X \cdot P|^{-\tilde{\Delta}, \epsilon}|2X' \cdot P|^{-\Delta, \epsilon}.
\end{split}
\end{equation}
Here, we have 
\begin{equation}
\label{Ddel}
D = \{(\Delta \in \mathbb{Z}, \epsilon) : \epsilon = \Delta \mod\mathbb{Z}_2\}. 
\end{equation}
The Plancherel measure $\frac{1}{\mu(\Delta)}$ is given by 
\begin{equation}
    \frac{1}{\mu(\Delta)}
    %=a(\rho)^2
    %=\frac{1}{2^2}~ 2^{1-3} ~\pi^{1-3} \tan{\frac{\pi}{2}\left(\Delta-1\right)}\tan{\frac{\pi}{2}\left(\tilde{\Delta}-1\right)}
    =\frac{1}{16 \pi^2}  \tan{\frac{\pi}{2}\left(\Delta-1\right)}\tan{\frac{\pi}{2}\left(\tilde{\Delta}-1\right)}.
\end{equation}
%\begin{equation}
%\label{deltahiden}
%\begin{split}
%\delta(X, X') &= \frac{1}{2\pi i} \sum_{\epsilon=0,1} \int_\Gamma \mathrm{d}\Delta \frac{\tan\left(\frac{\pi}{2}(\Delta-1)\right)\tan\left(\frac{\pi}{2}(\tilde{\Delta}-1)\right)}{32\pi^2} \int_{\mathrm{PC}_{2,2}} dP |2X \cdot P|^{-\tilde{\Delta}, \epsilon}|2X' \cdot P|^{-\Delta, \epsilon} \\
%&+ \sum_{\substack{(\Delta, \epsilon) \in D_{\text{}}}} \text{Res} \int_{\mathrm{PC}_{2,2}} \mathrm{d}P \frac{\tan\left(\frac{\pi}{2}(\Delta-1)\right)\tan\left(\frac{\pi}{2}(\tilde{\Delta}-1)\right)}{32\pi^2} |2X \cdot P|^{-\tilde{\Delta}, \epsilon}|2X' \cdot P|^{-\Delta, \epsilon}.
%\end{split}
%\end{equation}
Here, the shadow weight $\tilde{\Delta}=2-\Delta$.
We derive the eq.\eqref{Ddel} following \cite{Rossmann} noting that for $\mathrm{H}_{p, q}$ space
\begin{equation}
\begin{split}
\square f&=(\nu^2-\rho^2)f ,\quad \nu \in \mathbb{C}\\
\rho&=\frac{1}{2}(p+q-2)\\
\epsilon&=\rho+\nu-q ~\text{mod}~2.
\end{split}
\end{equation}
Now, looking at the Sturm-Liouville equation \eqref{strum} (for details we refer to appendix \ref{sleqnn}) and comparing it with $\square f=(\nu^2-\rho^2)f$ of \cite{Rossmann}  we get
\begin{equation}
\square f=(\nu^2-\rho^2)f=\Delta (\Delta-2)f.
\end{equation}
Now, $\rho=1$, for $p=q=2$, this gives
\begin{equation}
\begin{split}
\epsilon&=\rho+\nu-q ~\text{mod}~2\\
&=\Delta-2 ~\text{mod}~2\\
&\equiv\tilde{\Delta}~\text{mod}~2.
\end{split}
\end{equation}
By shadow symmetry we can also write the following condition for discrete part
\begin{equation}
\begin{split}
\epsilon&=\rho+\nu-q ~\text{mod}~2\\
&=\tilde{\Delta}-2 ~\text{mod}~2\\
&\equiv{\Delta}~\text{mod}~2.
\end{split}
\end{equation}
The contour $\Gamma$ represents half of the principal series \footnote{In\cite{Pasterski:2017kqt}, it was shown that the spectrum of conformal primary wavefunctions, characterized by principal series $\Delta \in 1 + i\mathbb{R}$, forms a complete set of delta-function-normalizable solutions, which are normalizable with respect to the Klein-Gordon inner product, for the massless Klein-Gordon equation. Similarly to the massless case, in \cite{Pasterski:2017kqt} it was shown that the spectrum of massive conformal primary wavefunctions, with principal series $\Delta \in 1 + i \mathbb{R}_{\geq0}$, constitutes a complete set of normalizable solutions to the massive Klein-Gordon equation.}, i.e. 
\begin{equation}
\int_{\Gamma}[~~]=\int_{1}^{1+i\infty}[~~].
\end{equation}

%$\mathrm{PC}_{2,2}$:

\section{Spectral representation in Klein space}
\label{sr}
Now, in this section \ref{sr}, we study the spectral representation in Klein space. The first step is to split the integral over Klein space into two separate integrals. One integral spans the timelike wedge $\mathcal{W}^T$, where $X^2<0$, while the other spans the spacelike wedge $\mathcal{W}^S$, where $X^2<0$. Within both the timelike and spacelike regions, the integral $\mathrm{d}^4X$ can be decomposed into two separate integrals. One integral ranges over $\tau=\sqrt{|X^2|}$, with $\tau$ spanning from $0$ to $\infty$. The other integral spans a single unit slice of $\mathrm{AdS}_3/\mathbb{Z}$. The measure yields
\begin{equation}
\begin{split}
\int_{\mathbb{R}^{2,2}} \mathrm{d}^4X=\int_{0}^{\infty} \mathrm{d}\tau \tau^3 \times \Bigg\{ \int_{\hat{x}_+^2=-1} \mathrm{d}^3 \hat{x}_++\int_{\hat{x}_-^2=1} \mathrm{d}^3 \hat x_- \Bigg\}.
\end{split}
\end{equation}
Now, we will move to the Feynman rules for celestial amplitudes in Klein space in next subsection \ref{Frulekl}.
\subsection{Feynman rules for celestial amplitudes in Klein space}
\label{Frulekl}
In this subsection \ref{Frulekl}, we explore the Feynman rules for celestial amplitudes in Klein space. Celestial amplitudes are scattering amplitudes expressed using conformal primary wavefunctions \cite{Pasterski:2016qvg,Pasterski:2017kqt}, instead of the standard plane-wave basis. For instance, to convert an $n \to m$ amputated amplitude for scalars in position space, $\mathcal{M}^{kl}(X_i)$, into a celestial scalar amplitude, $\mathcal{A}(x_i)$, we perform a basis transformation through an integral \eg see \cite{Chang:2023ttm, Liu:2024vmx, Liu:2024lbs}
\begin{equation}
\begin{split}
\mathcal{A}(x_i)=\bigg(\prod_{i=1}^{n+m}\int_{\mathbb{R}^{2,2}} \mathrm{d}^4 X_i\bigg)\bigg(\prod_{i=1}^{n}\phi_{\Delta_i}^{+}(x_i;X_i)\bigg)\bigg(\prod_{i=n+1}^{n+m}\phi^{-}_{\Delta_i}(x_i;X_{i})\bigg)\mathcal{M}^{kl}(X_i)\;.
\end{split}
\end{equation}
Here, $\phi^{\pm}_{\Delta}(x;X)$ represents the scalar conformal primary wavefunctions, with the superscripts $(+)$ and $(-)$ indicating incoming and outgoing conformal primary wavefunctions, respectively, and they possess a conformal dimension $\Delta$. These wavefunctions link a point $X$ in 4-dimensional Klein space to a point $x$ on the Lorentzian torus, $\mathcal{C}\mathcal{T}^2=S^1\times S^1$, for each slice of $\mathrm{AdS}_3/\mathbb{Z}$, which are the foliations of Klein space. This is why they are also known as bulk-to-boundary propagators. Next, we will examine the explicit forms of these conformal primary wavefunctions (bulk-to-boundary propagators) in Klein space.
In perturbation theory, the amputated amplitude in Klein space, $\mathcal{M}^{kl}(X_j)$, can be calculated using Feynman rules. The propagators and interaction vertices are derived by performing Fourier transforms on their momentum space equivalents. For instance, a scalar bulk-to-boundary propagator in Klein space that links two bulk points, $X_1$ and $X_2$, is expressed as follows \footnote{We use the Klein-space metric, \textit{i.e.} $\text{diag}(-1,-1,+1,+1)$.}
\begin{align}
\mathcal{K}_{m}^{kl}(X_1,X_2)=\int_{\mathbb{R}^{2,2}} \frac{\mathrm{d}^4p}{(2\pi)^{4}}\frac{i}{p^2-m^2}e^{-ip\cdot(X_1-X_2)}\;.
\end{align}
The equation for the Green's function of the bulk-to-bulk propagator is given by
\begin{align}
(\partial_X^2+m^2)\mathcal{K}_{m}^{kl}(X_1,X_2)=-i\delta^{(4)}(X_1-X_2)\;.
\end{align}
To check the bulk-to-boundary propagator in Klein space let us consider the Green's function of the
Klein-Gordon equation, i.e. some function $\mathcal{K}_{m}^{kl}(X_1,X_2)$ satisfying
\begin{equation}
(\partial_X^2+m^2)\mathcal{K}_{m}^{kl}(X_1,X_2)=-i\delta^{(4)}(X_1-X_2). 
\end{equation}
Fourier transforming this function gives 
\begin{equation}
\mathcal{K}_{m}^{kl}(X_1,X_2)= \int_{\mathbb{R}^{2,2}} \frac{\mathrm{d}^4p}{(2\pi)^{4}} e^{-ip\cdot(X_1-X_2)} \times \widetilde{\mathcal{K}}_{m}^{kl}(p). 
\end{equation}
In the $4$-momentum space, this becomes
\begin{equation}
(-p^2 + m^2) \times \widetilde{\mathcal{K}}_{m}^{kl}(p)= -i, 
\end{equation}
and therefore we get
\begin{equation}
\widetilde{\mathcal{K}}_{m}^{kl}(p)= \frac{i}{p^2 - m^2},
\end{equation}
and therefore we get 
\begin{equation}
\mathcal{K}_{m}^{kl}(X_1,X_2)=\int_{\mathbb{R}^{2,2}} \frac{\mathrm{d}^4p}{(2\pi)^{4}}\frac{i}{p^2-m^2}e^{-ip\cdot(X_1-X_2)}.
\end{equation}

The tree-level celestial amplitude in the $s$-channel involves two incoming and two outgoing massless scalars, with a massive scalar being exchanged. By utilizing the bulk-to-bulk and bulk-to-boundary propagators, we can compute the celestial amplitude for the $2 \to 2$ scattering process at tree level in the $s$-channel in Klein space \footnote{We have introduced the notation $i^{s}_m$ to label the $i$-th particle with mass $m$ and helicity/spin $s$ in the celestial amplitude.}.
\begin{align}
\hspace{-.45cm}{}_s\mathcal{A}^{\Delta_i}_{1^0_0+2^0_0\to 3^0_0+4^0_0}(x_i)=\int_{\mathbb{R}^{2,2}}  \mathrm{d}^4X_1\mathrm{d}^4X_2\bigg(\prod_{i=1}^2\phi^{+}_{\Delta_i}(x_i,X_1)\bigg)\mathcal{K}_{m}^{kl}(X_1,X_2)\bigg(\prod_{i=3}^4\phi^{-}_{\Delta_i}(x_i,X_2)\bigg)\;.
\end{align}
Likewise, the celestial scalar amplitude in Klein
space for the $2\to 2$ scattering process at tree level in the $t$-channel is
\begin{equation}
\begin{split}
{}_t\mathcal{A}^{\Delta_i}_{1^0_0+2^0_0\to 3^0_0+4^0_0}(x_i)=&\int_{\mathbb{R}^{2,2}}  \mathrm{d}^4X_1\mathrm{d}^4X_2\bigg(\prod_{i=1}^2\phi^{+}_{\Delta_i}(x_i,X_1)\bigg)\mathcal{K}_{m}^{kl}(X_1,X_2)\bigg(\prod_{i=3}^4\phi^{-}_{\Delta_i}(x_i,X_2)\bigg)\;.
\end{split}
\end{equation}
The expressions for the conformal primary wavefunctions $\phi^{\pm}_{\Delta,m}(x;X)$ for massive scalars with mass $m$ and $\phi^{\pm}_{\Delta}(x;X)$ for massless scalars are given by
\begin{equation}
\label{scalar_cpw}
\begin{split}
\phi^{\pm}_{\Delta,m}(x;X)&=\int\frac{\mathrm{d}^3\hat{p}^{\prime}}{\hat{p}^{\prime0}} \frac{1}{(-\hat{q}\cdot\hat{p}^{\prime})^{\Delta}} e^{\pm im\hat{p}^{\prime}\cdot X}\;,
\\
\phi^{\pm}_{\Delta}(x;X)&=\int_{0}^{\infty}\mathrm{d}\omega\,\omega^{\Delta-1}e^{\pm i\omega\hat{q}\cdot X-\varepsilon\omega}
=N^{\pm}_\Delta\frac{1}{(-\hat{q}\cdot X\mp i\varepsilon)^{\Delta}}\;,
\end{split}
\end{equation}
where $N^{\pm}_\Delta$ is a constant factor defined as
\begin{equation}
N^{\pm}_\Delta=(\mp i)^{\Delta}\Gamma[\Delta]\,.
\end{equation}
Now, $q^{\mu}$ and $p^{\mu}$ represent on-shell momenta for massless and massive particles in Klein space, respectively, satisfying $$-(q^0)^2-(q^1)^2+(q^2)^2+(q^3)^2=0,$$ and $$-(p^0)^2-(p^1)^2+(p^2)^2+(p^3)^2=-m^2,$$
with the following parameterizations
\begin{equation}
\label{klon}
\begin{split}
q^{\mu}&=\omega\hat{q}^{\mu}=\omega\left(1 + z\bar{z}, i(z + \bar{z}), i\left(\bar{z} - z\right), 1 - z\bar{z}\right)\,,
\\
p^{\mu}&= m\hat{p}^{\mu}=\frac{1}{2y}\left(1 + y^2 + z\bar{z}, i(z + \bar{z}), i\left(\bar{z} - z\right), 1 - y^2 - z\bar{z}\right).
\end{split}
\end{equation}
It is worth noting that in eq.\eqref{klon}, we use the Klein-space metric, \textit{i.e.} $\text{diag}(-1,-1,+1,+1)$ to parametrize the on-shell momenta for massless and massive particles. The points \( (z, \bar{z}) \) represent locations on the celestial torus where the massless particles enter, and exit. Also, we note that \((z, \bar{z})\) represent independent real variables within Klein space. This parametrization is related to the Lorentzian flat-space parametrization of on-shell momenta \eg see \cite{Pasterski:2016qvg} via analytic continuation. We refer to subsection \ref{anacon} for the details of the analytic continuation. 
For massless particles in Klein space have the momenta are given by  
\begin{equation}
\begin{split}
q_i^\mu &= \varepsilon_i \omega_i \hat{q}_i^\mu.
\end{split}
\end{equation}
Here, $i$ represents a particle label, $\omega_i \in (0, \infty)$ indicates the frequency magnitude, $\varepsilon_i \in \{ \pm 1\}$ signifies the frequency's sign, and $\hat{q}_i^\mu$ are null vectors defined using planar coordinates as in eq.\eqref{klon}.
%\begin{equation}
%\begin{split}
%\hat{q}_i^\mu &= \left(1+z_i \bar{z}_i, i(z_i+\bar{z}_i), i(z_i-%z_i \bar{z}_i), 1+z_i \bar{z}_i\right).
%\end{split}
%\end{equation}

An alternative set of conformal primary wavefunctions is obtained by applying the shadow transform to the conformal primary wavefunctions defined in eq.\eqref{scalar_cpw}, as shown in \cite{Pasterski:2017kqt}. The shadow transform of $\phi^{\pm}_{\Delta}(x;X)$ is given by
\begin{equation}
\begin{split}
    \widetilde{\phi}_{\Delta}^{\pm}(x;X)
     =\frac{\Gamma[2-\Delta]}{\pi\Gamma[1-\Delta]}\int \mathrm{d}^2x^{\prime}\frac{\phi_{2-\Delta}^{\pm}(x^{\prime};X)}{|x-x^{\prime}|^{2\Delta}}
    =\frac{N^\pm_{2-\Delta}}{N^\pm_{\Delta}}(-X^2)^{\Delta-1}\phi_{\Delta}^{\pm}(x;X)\;.
\end{split}
\end{equation}

At this stage, we first need to understand how to handle spectral representation in Klein space.
In the spectral representation studied in Euclidean AdS \cite{Penedones:2010ue, Costa:2014kfa, Meltzer:2019nbs} translated to celestial amplitude, the bulk-to-bulk propagator decomposes into bulk-to-boundary propagators \cite{Chang:2023ttm, Liu:2024vmx}. %We will prove this in section \ref{sprep}. 
Here, in Klein space the identity resolution on $\mathrm{AdS}_3/\mathbb{Z}$ has both continuous and discrete parts.
We introduce the conformal primary wavefunctions in Klein space  for massive, and tachyonic scalars %\footnote{When scattering amplitudes are calculated in a split signature $4d$ bulk, the celestial sphere takes on the form of a Lorentzian torus.} 
as
\begin{equation}
\label{sclbkl}
\begin{split}
\phi_{\Delta,m_{\text{R}},\epsilon}(x;X)&=\int\frac{\mathrm{d}^3\hat{p}}{|\hat{p}^{0}|}\frac{1}{|\hat{q}\cdot \hat{p}|^{\Delta}}\text{sgn}^{\epsilon}(\hat{q}\cdot \hat{p})e^{- im_{\text{R}}\hat{p}\cdot X},\\
\phi_{\Delta,im_{\text{I}},\epsilon}(x;X)&=\int\frac{\mathrm{d}^3\hat{p}}{|\hat{p}^{\star}|}\frac{1}{|\hat{q}\cdot \hat{p}|^{\Delta}}\text{sgn}^{\epsilon}(\hat{q}\cdot \hat{p})e^{- im_{\text{I}}\hat{p}\cdot X}.
\end{split}
\end{equation}
Here, for the massive conformal primary wavefunction $p^{2}=-m_{\text{R}}^2$, and for the tachyonic conformal primary wavefunction $p^{2}=m_{\text{I}}^2$, and $\hat{p}^{\star}=\hat{p}^{0}+\hat{p}^{ 3}$, and for both case $\epsilon=0, 1$. 
Now, we note that due to the identity resolution on $\mathrm{AdS}_3/\mathbb{Z}$, which includes both discrete and continuous parts, a new type of conformal primary wavefunction for the massive (or tachyonic) case, labeled by $\epsilon$, naturally arises as $\phi_{\Delta,m_{\text{R}},\epsilon}(x;X)$, and $\phi_{\Delta,im_{\text{I}}\epsilon}(x;X)$. 

We note that this particular form of the tachyonic conformal primary wavefunction was initially introduced in the CLM prescription \cite{Chang:2023ttm}, and has subsequently been utilized in works \cite{Liu:2024vmx, Liu:2024lbs}. The CLM prescription focuses on harmonic analysis within $\mathrm{dS}_{d+1}$, where they consider Minkowski space as being divided into regions foliated by past or future light cones (Euclidean AdS slices) and regions lying outside these cones (dS slices). In the resolution of identity within the dS slices, there exists both a continuous and a discrete parts. Thus, the $\epsilon$ label in the conformal primary wavefunction is only relevant for the spacelike momentum region, i.e. outside the light cone corresponding to the tachyonic scenario in the dS foliations of Minkowski space.
%We note that this form of the tachyonic conformal primary wavefunction appear first in  the CLM precription \cite{Chang:2023ttm}, and also used in \cite{Liu:2024vmx, Liu:2024lbs}. The CLM prescription deals with harmonic analysis on $\mathrm{dS}_{d+1}$ where they considered the fact that Minkowski space is foliated by regions having past or future light cones (Euclidean AdS slices) and regions outside these cones (dS slices). The resolution of identity in the dS slices has both continuous, and discrete part, so $\epsilon$ label in the conformal primary wavefunction appears only for the spacelike momentum region which corresponds to the tachyonic case in dS foliations of Minkowski space. %This represents a new conformal primary wavefunction that has appeared in Klein space. 

The tachyonic conformal primary wavefunction $\phi_{\Delta,im_{\text{I}},\epsilon}(x;X)$ satisfies the Klein-Gordon equation given by  
\begin{equation}
\begin{split}
(\partial^{2}_{X}+m_{\text{I}}^{2})\phi_{\Delta,im_{\text{I}},\epsilon}(x;X)=0.
\end{split}
\end{equation}

\subsection{Spectral representation for the bulk-to-bulk propagator in Klein-space}
\label{sprep}
In this section \ref{sprep}, we derive the spectral representation for the bulk-to-bulk propagator in Klein space $\mathcal{K}_{m}^{kl}(X_1,X_2)$ given by 
\begin{align}
&\mathcal{K}_{m}^{kl}(X_1,X_2)=\int_{\mathbb{R}^{2,2}} \frac{\mathrm{d}^{4}p}{(2\pi)^{4}}\frac{i}{p^2-m^2}e^{-ip\cdot(X_1-X_2)}\;.
\end{align}
Klein space can be divided into a timelike wedge denoted as $\mathcal{W}^T$ where $p^2<0$, and a spacelike wedge denoted as $\mathcal{W}^S$ where $p^2>0$ as in the table.
\begin{table}[h]
\centering
\begin{tabular}{|c|c|c|}
\hline
Wedge & Expression & Conditions \\
\hline
$\mathcal{W}^T$ & $p^\mu = m_{\text{R}}\hat p_+^\mu$ & $\hat p_+^2=-1$ \\
$\mathcal{W}^S$ & $p^\mu = m_{\text{I}}\hat p_-^\mu$ & $\hat p_-^2=+1$ \\
\hline
\end{tabular}
\caption{Timelike and spacelike wedge}
\end{table}
The bulk-to-bulk propagator in Klein space is split into the regions timelike momentum: \(p^2<0\), and spacelike momentum: \(p^2>0\) given by 
\begin{align}
\mathcal{K}_{m}^{kl}(X_1,X_2)=\mathcal{K}_{m-\text{tlike}}^{kl}(X_1,X_2)+\mathcal{K}_{m-\text{slike}}^{kl}(X_1,X_2)\;.
\end{align}
Here, for timelike momentum region the bulk-to-boundary propagator $\mathcal{K}_{m-\text{tlike}}^{kl}$ and for spacelike momentum region the bulk-to-boundary $\mathcal{K}_{m-\text{slike}}^{kl}$ are given by
\begin{align}
&\mathcal{K}_{m-\text{tlike}}^{kl}(X_1,X_2)=\int_{p^2<0}\frac{\mathrm{d}^{4}p}{(2\pi)^{4}}\frac{i}{p^2-m^2}e^{-ip\cdot(X_1-X_2)}\;,\\
&\mathcal{K}_{m-\text{slike}}^{kl}(X_1,X_2)=\int_{p^2>0}\frac{\mathrm{d}^{4}p}{(2\pi)^{4}}\frac{i}{p^2-m^2}e^{-ip\cdot(X_1-X_2)}\;.
\end{align}
Now, for timelike momentum region we calculate the bulk-to-boundary propagator $\mathcal{K}_{m-\text{tlike}}^{kl}$ in Klein space. For timelike momentum, we define $p^{\mu}:= m_{\text{R}}\hat{p}^{\mu}$ with $\hat{p}^2=-1$ and $\hat{p}^0>0$. 
Therefore, we have 
\begin{align}
\begin{split}
\mathcal{K}_{m-\text{tlike}}^{kl}(X_1,X_2)
=&\int_{-\infty}^{\infty}\frac{\mathrm{d}m_{\text{R}}}{(2\pi)^{4}}\frac{i|m_{\text{R}}|^{3}}{-m_{\text{R}}^2-m^2}\int\frac{\mathrm{d}^{3}\hat{p}_1}{\hat{p}_1^0}\int\frac{\mathrm{d}^{3}\hat{p}_2}{\hat{p}_2^0}\hat{p}_2^0\delta^{(3)}(\hat{p}_1-\hat{p}_2)e^{-im_{\text{R}}\hat{p}\cdot(X_1-X_2)}\;,
\end{split}
\end{align}
where we introduced a delta function into the equation. Now, we use the resolution of identity, i.e. the completeness relation in $\mathrm{AdS}_3/\mathbb{Z}$ eq.\eqref{riden} we get 
\begin{align}
\begin{split}
&\hat{p}_2^0\delta^{(3)}(\hat{p}_1-\hat{p}_2)\\
&=\frac{1}{2\pi i}\int_{\Gamma}\frac{\mathrm{d}\Delta}{2\mu(\Delta)}\int \mathrm{d}^2x|\hat{p}\cdot\hat{q}|^{-\Delta,\epsilon}|\hat{p}^{\prime}\cdot\hat{q}|^{\Delta-2,\epsilon}+ 
\sum_{(\Delta,\epsilon)\in D}\text{Res}\frac{(-1)^\epsilon}{2\mu(\Delta)}\int \mathrm{d}^2x|\hat{p}\cdot\hat{q}|^{-\Delta,\epsilon}|\hat{p}^{\prime}\cdot\hat{q}|^{\Delta-2,\epsilon}\;.
\end{split}
\end{align} 
The bulk-to-boundary propagator for timelike momentum $\mathcal{K}_{m-\text{tlike}}^{kl}$  is expressed as 
\begin{equation}
\begin{split}
\mathcal{K}_{m-\text{tlike}}^{kl}
=&\frac{1}{2}\int_{0}^{\infty}\frac{\mathrm{d}m_{\text{R}}}{(2\pi)^{4}}\frac{im_{\text{R}}^{3}}{-m_{\text{R}}^2+m^2}\left(\frac{1}{2 \pi i} \sum_{\epsilon=0,1} \int_{\Gamma}\frac{(-1)^\epsilon \mathrm{d} \Delta}{\mu(\Delta)} \int \mathrm{d}^2 x \phi_{\Delta, m_{\text{R}}, \epsilon}\left(x ; X_1\right) \phi_{2-\Delta, m_{\text{R}}, \epsilon}\left(x ; X_2\right)\right. \\ & \left.+ \sum_{(\Delta, \epsilon) \in D} \operatorname{Res} \frac{(-1)^\epsilon}{\mu(\Delta)} \int \mathrm{d}^2 x \phi_{\Delta, m_{\text{R}}, \epsilon}\left(x ; X_1\right) \phi_{2-\Delta, m_{\text{R}}, \epsilon}\left(x ; X_2\right)\right).
\end{split}
\end{equation}
Here, $\phi_{\Delta,im_{\text{I}},\epsilon}$ are the massive conformal primary wavefunctions as defined in first subeq. of eq.\eqref{sclbkl}. 
For spacelike momentum, we can define $p:=m_{\text{I}}\hat{p}$ with $m_{\text{I}}>0$, $\hat{p}^2=+1$. 
The invariant measure is 
\begin{align}
\int_{\mathbb{R}^{2,2}}  \mathrm{d}^{4}p=\int_{0}^{\infty}\mathrm{d}m_{\text{I}}\;m_{\text{I}}^{3}\int\frac{\mathrm{d}^{3}\hat{p}}{|\hat{p}^{\star}|}\;,
\end{align}
where, we defined $\mathrm{d}^{3}\hat{p}$ as shorthand for $\mathrm{d}\hat{p}^{\star}\mathrm{d}\hat{p}^1 \mathrm{d}\hat{p}^2$, i.e. $\mathrm{d}^{3}\hat{p}\equiv \mathrm{d}\hat{p}^{\star}\mathrm{d}\hat{p}^1 \mathrm{d}\hat{p}^2$ and $\hat p^{\star}$ as $\hat p^0+\hat p^{3}$, i.e. $\hat p^{\star}=\hat p^0+\hat p^{3}$
\begin{align}
\begin{split}
\mathcal{K}_{m-\text{slike}}^{kl}
=&\int_{0}^{\infty}\frac{\mathrm{d}m_{\text{I}}}{(2\pi)^{4}}\frac{im_{\text{I}}^{3}}{m_{\text{I}}^2-m^2}\int\frac{\mathrm{d}^{3}\hat{p}_1}{|\hat{p}_1^{\star}|}\int\frac{\mathrm{d}^{3}\hat{p}_2}{|\hat{p}_2^{\star}|}|\hat{p}_2^{\star}|\delta(\hat{p}_1^{\star}-\hat{p}_2^{\star})\delta^{(2)}(\hat{p}_1-\hat{p}_2)e^{-im_{\text{I}}\hat{p}_1\cdot(X_1-X_2)}\;.
\end{split}
\end{align}
%We have 
%\begin{align}
%\begin{split}
%\int \frac{\mathrm{d}^{3} \widehat{p}}{\widehat{p}^0}=\int \mathrm{d}^{4} \widehat{p} ~2 \delta\left(\widehat{p}^2+1\right) \theta\left(\widehat{p}^0\right).
%\end{split}
%\end{align}
Now, using the resolution identity, i.e. the completeness relation in $\mathrm{AdS}_3/\mathbb{Z}$ eq.\eqref{riden} we get 
\begin{align}
\begin{split}
&|\hat{p^{\star}}|\delta(\hat{p}^{\star}-\hat{p}^{\prime \star})\delta^{(2)}(\hat{p}-\hat{p}^{\prime})\\
&\quad=\frac{1}{2\pi i}\int_{\Gamma}\frac{\mathrm{d}\Delta}{2\mu(\Delta)}\int \mathrm{d}^2x|\hat{p}\cdot\hat{q}|^{-\Delta,\epsilon}|\hat{p}^{\prime}\cdot\hat{q}|^{\Delta-2,\epsilon}+ 
\sum_{(\Delta,\epsilon)\in D}\text{Res}\frac{(-1)^\epsilon}{2\mu(\Delta)}\int \mathrm{d}^2x|\hat{p}\cdot\hat{q}|^{-\Delta,\epsilon}|\hat{p}^{\prime}\cdot\hat{q}|^{\Delta-2,\epsilon}\;.
\end{split}
\end{align}
Now, we can write the the bulk-to-bulk propagator $\mathcal{K}^{kl}_{m-\text{slike}}$ as
\begin{equation}
\begin{split}
\mathcal{K}_{m-\text{slike}}^{kl}= & \frac{1}{2} \int_0^{\infty} \frac{\mathrm{d} m_{\text{I}}}{(2 \pi)^{4}} \frac{i m_{\text{I}}^{3}}{m_{\text{I}}^2-m^2}\left(\frac{1}{2 \pi i} \sum_{\epsilon=0,1} \int_{\Gamma}\frac{(-1)^\epsilon \mathrm{d} \Delta}{\mu(\Delta)} \int \mathrm{d}^2 x \phi_{\Delta, i m_{\text{I}}, \epsilon}\left(x ; X_1\right) \phi_{2-\Delta, i m_{\text{I}}, \epsilon}\left(x ; X_2\right)\right. \\ & \left.+ \sum_{(\Delta, \epsilon) \in D} \operatorname{Res} \frac{(-1)^\epsilon}{\mu(\Delta)} \int \mathrm{d}^2 x \phi_{\Delta, i m_{\text{I}}, \epsilon}\left(x ; X_1\right) \phi_{2-\Delta, i m_{\text{I}}, \epsilon}\left(x ; X_2\right)\right).
\end{split}
\end{equation}
Here, $\phi_{\Delta,im_{\text{I}},\epsilon}$ are the tachyonic conformal primary wavefunctions as defined in second subeq. of eq.\eqref{sclbkl}. %We have
%\begin{equation}
%D = \{(\Delta \in \mathbb{Z}, \epsilon) : \epsilon = \Delta \mod \mathbb{Z}_2\}. 
%\end{equation}
The bulk-to-bulk propagator in Klein space $\mathcal{K}_m(X_1,X_2)$ can be represented as follows
\begin{equation}
\label{klspecvv}
\begin{split}
\mathcal{K}_{m}^{kl}(X_1,X_2)&=\mathcal{K}_{m-\text{tlike}}^{kl}(X_1,X_2)+\mathcal{K}_{m-\text{slike}}^{kl}(X_1,X_2)\\
&=\frac{1}{2}\int_{0}^{\infty}\frac{\mathrm{d}m_{\text{R}}}{(2\pi)^{4}}\frac{im_{\text{R}}^{3}}{-m_{\text{R}}^2+m^2}\left(\frac{1}{2 \pi i} \sum_{\epsilon=0,1} \int_{\Gamma}\frac{(-1)^\epsilon \mathrm{d} \Delta}{\mu(\Delta)} \int \mathrm{d}^2 x \phi_{\Delta, m_{\text{R}}, \epsilon}\left(x ; X_1\right) \phi_{2-\Delta, m_{\text{R}}, \epsilon}\left(x ; X_2\right)\right. \\ & \left.+ \sum_{(\Delta, \epsilon) \in D} \operatorname{Res} \frac{(-1)^\epsilon}{\mu(\Delta)} \int \mathrm{d}^2 x \phi_{\Delta, m_{\text{R}}, \epsilon}\left(x ; X_1\right) \phi_{2-\Delta, m_{\text{R}}, \epsilon}\left(x ; X_2\right)\right)\\
& +\frac{1}{2} \int_0^{\infty} \frac{\mathrm{d} m_{\text{I}}}{(2 \pi)^{4}} \frac{i m_{\text{I}}^{3}}{m_{\text{I}}^2-m^2}\left(\frac{1}{2 \pi i} \sum_{\epsilon=0,1} \int_{\Gamma}\frac{(-1)^\epsilon \mathrm{d} \Delta}{\mu(\Delta)} \int \mathrm{d}^2 x \phi_{\Delta, i m_{\text{I}}, \epsilon}\left(x ; X_1\right) \phi_{2-\Delta, i m_{\text{I}}, \epsilon}\left(x ; X_2\right)\right. \\ & \left.+ \sum_{(\Delta, \epsilon) \in D} \operatorname{Res} \frac{(-1)^\epsilon}{\mu(\Delta)} \int \mathrm{d}^2 x \phi_{\Delta, i m_{\text{I}}, \epsilon}\left(x ; X_1\right) \phi_{2-\Delta, i m_{\text{I}}, \epsilon}\left(x ; X_2\right)\right).
\end{split}
\end{equation}
Here, 
\begin{equation}
\begin{split}
   \frac{1}{\mu(\Delta)}
    &=\frac{1}{16 \pi^2}  \tan{\frac{\pi}{2}\left(\Delta-1\right)}\tan{\frac{\pi}{2}\left(\tilde{\Delta}-1\right)},\\
    \tilde{\Delta}&=2-\Delta,\\
    D &= \{(\Delta \in \mathbb{Z}, \epsilon) : \epsilon = \Delta \mod \mathbb{Z}_2\}. 
\end{split}
\end{equation}
%\textcolor{red}{add other details}
The eq.\eqref{klspecvv} is the spectral representation formula of the bulk-to-bulk propagator in Klein space. From the spectral representation formula of eq.\eqref{klspecvv}, we find that the bulk-to-bulk propagator in Klein space can be interpreted as a sum of products of two massive (or tachyonic) bulk-to-boundary propagators. The first two lines in eq.\eqref{klspecvv} represent the contribution from a massive scalar with mass \(m_{\text{R}}\), while the last two lines represent the contribution from a tachyonic scalar with imaginary mass \(im_{\text{I}}\). Both pairs of massive (or tachyonic) conformal primary wavefunctions have conformal dimensions \(\Delta\) and \(2-\Delta\), sharing a common boundary coordinate, with the conformal dimension \(\Delta\) and the real (or imaginary) mass integrated. The first and third lines in eq.\eqref{klspecvv} derive from the continuous part of the resolution of identity on \(\mathrm{AdS}_3/\mathbb{Z}\), whereas the second and fourth lines derive from the discrete part of the resolution of identity on \(\mathrm{AdS}_3/\mathbb{Z}\).
\section{Conclusions and future directions}
In this paper, we initiate the study the spectral representation in Klein space. We start with Klein space kinematics. We explain the hyperbolic foliation of Klein space in terms of Lorentzian $\mathrm{AdS}_3/\mathbb{Z}$ slices. We explain celestial amplitudes in Klein space and their building blocks called celestial leaf amplitudes. We explicitly show that the identity resolution on Lorentzian $\mathrm{AdS}_3/\mathbb{Z}$ has both continuous and discrete parts. Then, we use the foliations of Klein space into $\mathrm{AdS}_3/\mathbb{Z}$ slices, and the resolution of identity on $\mathrm{AdS}_3/\mathbb{Z}$ slices to give a proof-of-principle prescription of spectral representation in Klein space. We compute the Plancherel measure and resolution of identity in $\mathrm{AdS}_3/\mathbb{Z}$, and we also give a 
relation to the Plancherel measure of the CLM prescription for Euclidean AdS space and dS spacetime \cite{Chang:2023ttm} which gives a nice consistency check of our prescription. 

Using the resolution of identity in $\mathrm{AdS}_3/\mathbb{Z}$ having both continuous and discrete parts, we calculate the spectral representation of the massive bulk-to-bulk propagator propagator in Klein space. It can be expressed as the product of two  massive (or tachyonic) conformal primary wavefunctions having  of real (or imaginary) mass with both continuous and discrete parts, each with conformal dimensions $\Delta$ and $2-\Delta$, sharing a common boundary coordinate. Both the discrete and continuous parts involve integrating the common boundary coordinate and the real (or imaginary) mass. In the discrete part, the conformal dimension $\Delta$ is summed, whereas in the continuous part, it is integrated. A noteworthy observation is that when the identity resolution has both discrete and continuous parts, it results in a new conformal primary wavefunction for the massive (or tachyonic) \eg see eq.\eqref{sclbkl}. We note that this specific form of the tachyonic conformal primary wavefunction was initially introduced within the CLM prescription \cite{Chang:2023ttm}, and has since been employed in subsequent studies \eg see \cite{Liu:2024vmx, Liu:2024lbs}. In the CLM prescription on harmonic analysis within $\mathrm{dS}_{d+1}$, Minkowski space is divided into regions foliated by past or future light cones (Euclidean AdS slices) and regions exterior to these cones (dS slices). In the resolution of identity within the dS slices, both continuous and discrete parts are present. Consequently, the $\epsilon$ label in the conformal primary wavefunction pertains solely to the spacelike momentum region, which corresponds to the tachyonic scenario in the dS foliations of Minkowski space.

The spectral representation in Klein space is a computational tool to derive conformal block expansions for celestial amplitudes, along with their building blocks called celestial leaf amplitudes. This technique involves integrating the particle interaction point over a single layer of foliation. We need to use our prescription in various directions to make it a computational tool. We therefore end our paper by outlining several future directions.
%By breaking down the complex interactions into simpler components, it provides a clearer and more efficient way to analyze and understand the structure and symmetries of celestial amplitudes in Klein space. This approach is essential for simplifying the calculations and gaining deeper insights into the underlying physics of particle interactions.

\paragraph{Future directions.}
\paragraph{Four-point conformal block expansion in Klein space.}
%We can utilize the spectral representation in Klein space to analyze different celestial amplitudes and deduce the conformal partial wave expansions for them.
%We can use spectral representation to calculate the conformal partial wave expansion of a four-point contact amplitude. %We can consider the example of massless scalar four-point contact amplitude. 
Employing the spectral representation in Klein space, we compute the conformal block expansions for celestial amplitudes in Klein space for four-point tree-level massless, and massive exchanges. We compute the conformal block expansions for four-point celestial amplitudes of two massive and two massless scalars in the external legs with massive exchange. We will report more details on this in a forthcoming paper \cite{DMII}.
\paragraph{Adding legs.}
Using the spectral representation in Klein space, we can examine the conformal block expansions for celestial amplitudes in Klein space for five and six-point interactions with massless or massive scalar exchanges, as developed for Minkowski space in \cite{Liu:2024lbs}.
%Using the spectral representation in Klein space, we can analyze the conformal block expansions for the  celestial amplitudes in Klein space of five and six-point with massless/massive scalar exchanges \eg see \cite{Liu:2024lbs} developed for Minkowski space. 
In Minkowski space, the authors of \cite{Liu:2024lbs} concluded that using the standard conformal primary basis for all massless particles made it is impossible to close the contour in the conformal block expansion. To tackle this, they opted for the shadow conformal basis for outgoing particles. It would be great to understand whether their findings still hold true for celestial amplitudes in Klein space, or if they give new physics. At the six-point amplitudes, there exist two distinct topologies: the comb channel and the snowflake channel \eg see \cite{Liu:2024lbs, Fortin:2020yjz}. It would be great to analyze the celestial amplitudes in Klein space in these channels. We can assume that all external legs are massless scalars for both five-point and six-point amplitudes. Alternatively, we can consider scenarios with one massive scalar and four massless scalars as the external legs, and allow for both massless and massive exchanges.
%We can consider external legs to be all massless scalars for both five and six-point amplitude. We can consider also one massive and four massless scalars for external legs, and massless/massive exchange.

\paragraph{Adding spins.}
We explore the spectral representation for celestial amplitudes in Klein space to include spinning bulk-to-boundary propagators. %While our paper exclusively derived the spectral representation for spinless Feynman propagators in Klein
%space, it is crucial to establish similar representations for spinning Feynman propagators in order to calculate celestial amplitudes involving them. 
Using the spectral representations for spinning bulk-to-boundary propagators, we study the conformal block expansion of celestial amplitudes in Klein space for exchanged photons and gravitons. For that, we generalize the spinning AdS propagators prescription in Euclidean AdS \cite{Costa:2014kfa} to $\mathrm{AdS}_3/\mathbb{Z}$, i.e. foliations of Klein space. We will report more details on this in a forthcoming paper \cite{DMIV}.  
\paragraph{Adding loops.} Using the spectral representation in Klein space, we can give a systematic framework to calculate celestial loop amplitudes in Klein space. In momentum space, loop integrals often diverge due to bulk UV divergences and necessitate regularization. The spectral representation of bulk-to-bulk propagators in Klein space enables expressing a specific loop diagram using spectral integrals and combined products of higher-order tree diagrams. This simplifies loop calculations to assessing the spectral integrals mentioned, along with conformal integrals arising from the tree diagrams. We leave an understanding of celestial loop amplitudes in Klein space to future work.

\paragraph{Factorization singularities of massive amplitudes in Klein space.} Using the spectral representation in Klein space, we explore the factorization singularities of massive amplitudes in Klein space. The factorization singularity related to massive poles mandates that the kinematics of an $n$-point amplitude can be mapped into a configuration resembling a contact Witten diagram in $\mathrm{AdS}_3/\mathbb{Z}$. We will report more details on this in a forthcoming paper \cite{DMIII}.  

%\paragraph{Connection with 2d/0d dictionary.} It would be great to generalize the spectral represenation in Klein space to celestial amplitude in 2d/0d dictionary developed in \cite{Duary:2022onm, Kapec:2022xjw, Stolbova:2023smk, Stolbova:2023cof}. It is worth highlighting that when $p=q=1$, $H_{p,q}$ coincides with the $2d$ Minkowski space. In this direction, it would be nice to make connection with the techniques developed in gapped and gappless $2d$ QFTs in the flat-space limit of AdS, i.e. $\mathrm{QFT}_2/\mathrm{CFT}_1$ \cite{Cordova:2022pbl, Duary:2023gqg}. In particular, it would be nice to extract the celestial amplitude from the Polyakov blocks of $\mathrm{CFT}_1$ in the flat-space limit.
\paragraph{Carrollian amplitudes in Klein space.} By employing the spectral representation Klein space, we can calculate carrollian amplitudes in Klein space, which retains information about the null direction. We can then extend this to add legs, spins and loops in carrollian amplitudes.

\paragraph{Extrapolate definition of celestial amplitudes in Klein
space.}
Recently, Sleight and Taronna introduced a new approach \cite{Sleight:2023ojm} to define celestial amplitudes in Minkowski space. Their approach involves applying a Mellin transform to the radial direction within the hyperbolic slicing of Minkowski space, along with boundary limits in the hyperbolic directions. This approach aims to address an ambiguity in the conventional method for celestial amplitudes arising from a divergent integral in the definition of conformal primary wavefunctions. They showed that by defining perturbative celestial amplitudes in this way, one can link them to Witten diagrams in Euclidean AdS space. Using the spectral representation in Klein space, it would be great to develop a similar approach in Klein space to establish connections between celestial amplitudes in Klein space and Witten diagrams arising from the foliations of Klein space in Lorentzian $\mathrm{AdS}_3/\mathbb{Z}$.

\paragraph{Linking the string worldsheet with the Lorentzian CCFT.} 

In celestial amplitudes with massless external states, the Mellin transform integrates over all energy, eliminating the distinction between low (IR) and high energy (UV). After energy integrals, celestial string amplitudes are expressed as integrals over moduli, prompting the question of a limit based on celestial data that links to the celestial sphere. Such a limit exists \eg see \cite{Castiblanco:2024hnq}, identified by a large imaginary total scaling dimension. This finding suggests a correspondence between the string worldsheet, punctuated by vertex operator insertions, and the celestial sphere, punctuated by conformal primary operators representing the momentum of each external state. The connection is clearer for closed strings, particularly at tree level where the worldsheet's topology is a sphere. Exploring the spectral representation in Klein space, which corresponds to a Lorentzian torus for $4d$ bulk Lorentzian celestial amplitudes, could help understand the genus expansion of string theory from the perspective of Lorentzian CCFT.

\paragraph{Celestial multi-time-folds amplitudes in Klein space.}
It would be nice to develop a framework of celestial multi-time-folds amplitudes in Klein space from generalized amplitudes essentially determined as vacuum expectations of ``in'' and ``out'' creation/annihilation operators with arbitrary operator ordering, representing a relaxation of time-ordering constraints, \eg see \cite{Caron-Huot:2023vxl, Borsten:2024dvq}. 
%When using the standard path integral, we calculate time-ordered correlators. Conversely, anti-time-ordered correlators can be derived by reversing the path integral in time. Multi-time-fold amplitudes combine anti-time-ordered and time-ordered operators, linking their respective Lorentzian real-time axes through a small Euclidean imaginary time. This results in a path integral that oscillates between the past and future across multiple time-folds.
%When employing the standard path integral, we compute time-ordered correlators, while anti-time-ordered correlators can be obtained by running a path integral backward in time. Multi-time-folds amplitudes represent a combination of anti-time-ordered and time-ordered operators, connecting their respective Lorentzian real-time axes via small Euclidean imaginary time, resulting in a path integral oscillating between the past and future through multiple time-folds.
%\paragraph{Spinning Feynman propagators.}
\section*{Acknowledgements}
We would like to thank E. K. Narayanan for useful discussions. The work of S.D. is supported by the ICTS-Research Associate position of the TIFR Graduate School, and the Department of Atomic Energy, Government of India, under project no. RTI4001. The work of S.M. is supported by fellowship from CSIR, Govt. of India. 
\appendix
\section{Intermezzo: mathematical preliminaries}
\label{appA}
In this appendix \ref{appA}, we give an overview of the essential mathematical ideas related to spectral representation in Klein space. We refer to \cite{Strichartz1973HarmonicAO, Strichartz1989HarmonicAO, Heckman, Dobrev, Rossmann}, see \eg \cite{Chang:2023ttm}, and also \cite{Folland}. 
\subsection{Homogeneous spaces}
The homogeneous space $G / H$, where $H$ is a closed subgroup of $G$, can be understood as the collection of left cosets $g H$ for all elements $g$ in $G$, modulo an equivalence relation, i.e. $G / H=\{g H: g \in G\} / \sim$. In this context, the origin $e$ in $G / H$ corresponds to the coset of the identity $e H$, and $H$ acts as the stabilizer subgroup, also known as the ``little group,'' centered around this origin.

In the case of $\mathrm{AdS}_3/\mathbb{Z}$, the group $G$ takes the form of a direct product $G' \times G'$, and $H$ is identified as the diagonal subgroup, specifically $H$ is isomorphic to $G'$ like $H \simeq G^{\prime}$. For example
\begin{equation}
\mathrm{AdS}_3 / \mathbb{Z} \simeq \mathrm{SL}(2, \mathbb{R})_{\text{left}}\cross  \mathrm{SL}(2,\mathbb{R})_{\text{right}} /  \mathrm{SL}(2,\mathbb{R})_{\text{diag}}.
\end{equation}
Therefore, up to global identifications
\begin{equation}
\mathrm{AdS}_3 / \mathbb{Z} \simeq \mathrm{SL}(2, \mathbb{R})_{\text{left}}\cross  \mathrm{SL}(2,\mathbb{R})_{\text{right}}.
\end{equation}
From a group theory perspective, the real hyperbolic space $\mathrm{AdS}_3 / \mathbb{Z}$ can alternatively be expressed as
\begin{equation}
\mathrm{H}_{2, 2} = \frac{\mathrm{SO}(2, 2)}{\mathrm{SO}(2,1)} = G/H.
\end{equation}
\subsection{$H$-fixed vector and $H$-fixed distribution}
The averaging operator 
\begin{equation}
\mathcal{P}_H \cdot f(x)=\int_H d h f(x h),
\end{equation}
defined through the right action of $H$ when $H$ is compact, acts as a projector from $L^2(G)$ to $L^2(G / H)$. Consequently, the set $\widetilde{G}^H$ of unitary irreducible representations in $L^2(G / H)$ is a proper subset of the tempered unitary dual $\widetilde{G}^0$ of $L^2(G)$.

In our context, the averaging projector $\mathcal{P}_H$ becomes problematic because $H$ is non-compact, causing the normalizable functions in $L^2(G / H)$ to lose their normalizability in $L^2(G)$. This leads to the possibility of additional irreducible representations in $\widetilde{G}^H$ that do not exist in $\widetilde{G}^0$, and the decomposition of $L^2(G / H)$ into irreducible components may not directly relate to that of $L^2(G)$. We can characterize these extra irreps in $\widetilde{G}^H$ by considering $H$-fixed vectors as $H$-fixed distributions.

In the $\mathrm{AdS}_3 / \mathbb{Z}$ spacetime, the bulk-to-boundary covariant quantity $X.P$ is indefinite, enabling the presence of two distinct spherical functions associated with the same unitary irreducible representation, where $\Delta=1+ i s, s \in \mathbb{R}$.

\subsection{$H$-spherical function}
The $H$-left-invariant functions on $G / H$ are characterized by the condition 
\begin{equation}
f\left(h^{-1} x\right)=f(x),
\end{equation}
for $x \in G / H$ and $h \in H$. Alternatively, they exhibit $H$-bi-invariance on $G$, expressed as $f\left(h_1^{-1} g h_2\right)=f(g)$ for $g \in G$ and $h_1, h_2 \in H$. For any $H$-fixed vector $v$, where $\pi(h) v=v$, a naturally corresponding $H$-bi-invariant function $\phi_v(g)$ is defined as
\begin{equation}
\phi_v(g) := (\pi(g) v, v), \quad \text{for } g \in G.
\end{equation}
This simplifies to an $H$-left-invariant function $\phi_v(x)$ on $G / H$ when $g=x h$. These functions, originating from $H$-fixed vectors, are termed $H$-spherical functions on $G / H$.

\subsection{Spherical function in $\mathrm{AdS}_3/\mathbb{Z}$}
We opt for a reference point $X_0=(0, \ldots, 0,1)$ in $\mathrm{AdS_3}/\mathbb{Z}$, where the stabilizer subgroup of $X_0$ is denoted by $H \simeq \mathrm{SO}(2, 1)$. In this scenario, the $\mathrm{SO}(2, 1)$-orbits are defined by $$P^{3}=\text{constant},$$ and the homogeneous functions are determined by their values on the two orbits $P^{3}= \pm 1$. Consequently, two $H$-fixed homogeneous functions exist on $\mathrm{PC}_{2,2}$
\begin{equation}
f_{\Delta, \epsilon}(P) = \left|P^{3}\right|^{-\Delta, \epsilon} := \left|P^{3}\right|^{-\Delta} \operatorname{sgn}^\epsilon\left(P^{3}\right), \text{ for } P \in \mathrm{PC}_{2,2} .
\end{equation}
The lightcone \(P \cdot P = 0\) for \(P \neq 0\) and \(P \in \mathbb{R}^{2,2}\) is denoted as \(\text{LC}_{2,2}\). The asymptotic boundary of \(\text{H}_{2,2}\) is the projective lightcone \(\text{PC}_{2,2}\), defined by
\[
\text{PC}_{2,2} := \text{LC}_{2,2} / \mathbb{R}^+ \cong S^1 \times S^1.
\]
The superscript notation $\epsilon=0,1$ denotes the parity of $f_{\Delta, \epsilon}(P)$ with respect to the transformation $P \rightarrow -P$. The matrix element is given by $\pi(X) f_{\Delta, \epsilon}(P)=|X \cdot P|^{-\Delta, \epsilon}$.

The functions $f_{\Delta, \epsilon}(P)$ exhibit singularities on the hypersurface $P^3=0$, and we can associate $H$-fixed distributions with $\mathcal{E}_{\Delta}$. For the principal series $\mathcal{E}_{1+i s}$, where $s \in \mathbb{R}$, the corresponding spherical functions are
\begin{equation}
\phi_{\Delta, \epsilon}(T) = \phi_{\Delta, \epsilon}\left(X_1, X_2\right) = \int_{\mathrm{PC}_{2,2}} \mathrm{d}P \left|X_1 \cdot P\right|^{-\bar{\Delta}, \epsilon} \left|X_2 \cdot P\right|^{-\Delta, \epsilon}.
\end{equation}
Here, $T=X_1 \cdot X_2$, and they adhere to both parity symmetry and shadow symmetry
\begin{equation}
\phi_{\Delta, \epsilon}(T) = (-1)^\epsilon \phi_{\Delta, \epsilon}(-T) = \phi_{\tilde{\Delta}, \epsilon}(T) .
\end{equation}
\subsection{Poisson and Fourier transform}
For every $H$-fixed vector $v$ within the irreducible representation $\mathscr{H}_\pi$ with the inner product denoted as $(\cdot, \cdot)$, an intertwining operator $\mathscr{F}_v^{\dagger}$ can be formulated from $\mathscr{H}_\pi$ to $L^2(G / H)$
\begin{equation}
\mathscr{F}_v^{\dagger}: w \in \mathscr{H}_\pi \mapsto (\pi(x) v, w), \quad \text{for } x \in G / H .
\end{equation}
The representation of $\mathscr{H}_\pi$ in $L^2(G / H)$ should be perceived as a part of the direct integral. Through the Schur lemma and the irreducibility of $\pi$, $\mathscr{F}_v^{\dagger}$ acts as an embedding from $\mathscr{H}_\pi$ into $L^2(G / H)$, also referred to as the Poisson transform. Defined by $$\left(\mathscr{F}_v^{\dagger}[w], f(x)\right)=\left(w, \mathscr{F}_v[f(x)]\right),$$ the adjoint of the Poisson transform $\mathscr{F}_v^{\dagger}$ is known as the smearing operator $\mathscr{F}_v$
\begin{equation}
\mathscr{F}_v: f(x) \in L^2(G / H) \mapsto \int_{G / H} d x \, f(x) \, \pi(x) v.
\end{equation}
This is also referred to as the Fourier transform of $f(x)$. Subsequently, the projector $$\mathcal{P}_v:=\mathscr{F}_v^{\dagger} \mathscr{F}_v$$ isolates the $\mathscr{H}_\pi$ component of $f(x) \in L^2(G / H)$
\begin{equation}
\mathcal{P}_v: f(x) \mapsto \int_{G / H} d y \, f(y) \, (\pi(x) v, \pi(y) v) = \int_{G / H} d y \, f(y) \, \phi_v\left(y^{-1} x\right) .
\end{equation}
In the last equality, we have employed the principle that when $f_1 \in L^2(G / H)$ and $f_2$ is spherical, the convolution on $G / H$ is well-defined
\begin{equation}
\left(f_1 * f_2\right)(x) := \int_{G / H} d y \, f_1(y) \, f_2\left(y^{-1} x\right), \quad \text{for } f_1, f_2 \in L^2(G / H) .
\end{equation}
The Poisson transform translates the $H$-fixed vector $v$ into the $H$-spherical function $\phi_v(x)$.

\subsection{Completeness relation}
In the case of compact and Riemannian semisimple symmetric spaces, the collection of spherical functions $\phi_v(x)$ linked with $H$-fixed vectors forms a complete orthogonal basis for $H$-left-invariant functions on $G / H$. When considering the Dirac delta distribution $\delta(x)$ on $G / H$, we observe that $\mathscr{P}_v[\delta(x)]=\phi_v(x)$, and the completeness relationship can be expressed as
\begin{equation}
\delta(x) = \int_{\tilde{G}^{H, c}} \left|c^c(v)\right|^{-2} d v \, \phi_v(x) + \sum_{\tilde{G}^{H, d}} \left|c^d(v)\right|^{-2} \, \phi_v(x).
\end{equation}
Here, $\widetilde{G}^{H, c}$ and $\widetilde{G}^{H, d}$ denote the continuous and discrete parts, respectively, of $\widetilde{G}^H$ within the decomposition of $L^2(G / H)$. The density can be determined as
\begin{equation}
\begin{split}
\left|c^c(v)\right|^2 \delta(u, v) & = (v, v)^{-1} (\phi_u, \phi_v) \\
\left|c^d(v)\right|^2 \delta_{u, v} & = (v, v)^{-1} (\phi_u, \phi_v).
\end{split}
\end{equation}
Here, the inner product of spherical functions follows the structure of $L^2(G / H)$. The inversion formula for $f(x) \in L^2(G / H)$ is given by
\begin{equation}
\begin{split}
f(x) &= \int_{\widetilde{G}^{H, c}} \left|c^c(v)\right|^{-2} d v \, \mathscr{P}_v[f(x)] \quad + \sum_{\widetilde{G}^{H, d}} \left|c^d(v)\right|^{-2} \, \mathscr{P}_v[f(x)].
\end{split}
\end{equation}

In the literature, $c(v)$ and $|c(v)|^{-2}$ are commonly referred to as the Harish-Chandra $c$-function and the Plancherel measure of $G / H$, respectively.

\subsection{Resolution of identity on $\mathrm{H}_{p, q}$ into continuous and discrete parts}
\label{discont}
In this subappendix \ref{discont}, we explore the resolution of identity on $\mathrm{H}_{p, q}$ into continuous and discrete parts. 
The hyperbolic space of our interest is given by  
\begin{equation}
\mathrm{H}_{p, q}=\mathrm{SO}(p, q) / \mathrm{SO}(p, q-1).
\end{equation}
First, let us consider the problem of resolution of identity on $\mathrm{H}_{p, q}$ into continuous and discrete parts in a broader context. If we assume \( G \) is a non-compact semisimple Lie group, and \( H \) is a closed subgroup, then \( G/H \) forms what we call a homogeneous space. This space may or may not possess a \( G \)-invariant measure, \eg see \cite{Folland}.

A general question one explores in this setting is how the \( G \)-action on \( L^2(G/H) \) decomposes into irreducible representations of \( G \).

The answer, naturally, varies. Here, we will mention two specific scenarios: 
\begin{enumerate}[label=\roman*.]

\item If \( H = K \), where \( K \) is a maximal compact subgroup of \( G \), then \( G/K \) has only a continuous spectrum. Conversely, if \( H = \Gamma \), where \( \Gamma \) is a discrete subgroup such that \( G/\Gamma \) is compact, we observe only a discrete spectrum. Generally, one may encounter both continuous and discrete spectra. %For instance, in the case of \( \mathrm{SL}(2, \mathbb{R}) \)/\( \mathrm{SL}(2, \mathbb{Z}) \), there are both continuous and discrete spectra.

\item In the space of interest, namely, \( \mathrm{SO}(p, q) / \mathrm{SO}(p, q-1) \), following \cite{Strichartz1973HarmonicAO}, we can show that for \( q > 1 \), the spectrum has both continuous and discrete parts. In other words, if we take a suitable function, say in \( C_c^{\infty} \) of the space, the objective is to express it as the inverse Fourier transform involving the eigenfunctions of the Laplacian. If the spectrum contains both parts, then this decomposition will also have both continuous and discrete parts. 
In this context, \( C_c^{\infty} \) denotes the space of smooth functions with compact support. To elaborate

- \( C^{\infty} \) represents the set of all infinitely differentiable (smooth) functions.\\
- \( C_c \) signifies that these functions have compact support, meaning each function is zero outside a bounded subset of the domain.

Thus, \( C_c^{\infty} \) comprises all smooth functions that vanish outside some compact region of the space.

\end{enumerate}

\section{Laplacian in $\mathrm{AdS}_3/\mathbb{Z}$ global coordinate and the Sturm-Liouville equation}
\label{sleqnn}
In this appendix \ref{sleqnn}, we study the laplacian in $\mathrm{AdS}_3/\mathbb{Z}$ global coordinate and the Sturm-Liouville equation.
The unit metric on $\mathrm{AdS}_3/\mathbb{Z}$ in the global coordinates is given by
\begin{equation}
	\mathrm{d} s_3^2=-\cosh^2\rho \mathrm{d} \psi^2+\mathrm{d} \rho^2+\sinh^2\rho \mathrm{d} \phi^2.
\end{equation}
The $\mathbb{Z}$ acts as the time-like quotient $\psi \sim \psi+2\pi$.
The laplacian eigenvalue equation (Sturm-Liouville equation) is given by 
\begin{equation}\label{strum}
\begin{split}
\square f-m^2f&=0\\
\implies \square f-\Delta (\Delta-2)f&=0.
\end{split}
\end{equation}
The laplacian is 
\begin{equation}
\begin{split}
\square f
%&=\frac{1}{\sqrt{-\text{detg}}}\partial_{\mu}\Big(\sqrt{-\text{detg}}~g^{\mu \nu}\partial_{\nu}f\Big)\\
%&=\frac{1}{\sqrt{-\text{detg}}}\partial_{\psi}\Big(\sqrt{-\text{detg}}~g^{\psi \psi}\partial_{\psi}f\Big)+\frac{1}{\sqrt{-\text{detg}}}\partial_{\rho}\Big(\sqrt{-\text{detg}}~g^{\rho \rho}\partial_{\rho}f\Big)\\
%&+\frac{1}{\sqrt{-\text{detg}}}\partial_{\phi}\Big(\sqrt{-\text{detg}}~g^{\phi \phi}\partial_{\phi}f\Big)\\
%&=\frac{1}{\sinh \rho \cosh \rho}\partial_\psi (\sinh \rho \cosh \rho \times -\sech^2\rho \partial_{\psi}f)+\frac{1}{\sinh \rho \cosh \rho}\partial_\rho (\sinh \rho \cosh \rho  \partial_{\rho}f)\\
%&+\frac{1}{\sinh \rho \cosh \rho}\partial_\phi (\sinh \rho \cosh \rho \times \csch^2\rho \partial_{\phi}f)\\
&=-\sech^2\rho \frac{\partial^2f}{\partial \psi^2}+\frac{\partial^2f}{\partial \rho^2}+(\tanh \rho +\coth \rho) \frac{\partial f}{\partial \rho}+\csch^2\rho \frac{\partial^2f}{\partial \phi^2}.
\end{split}
\end{equation}
%\textcolor{blue}{Global coordinates.}
In global coordinates we have 
\begin{equation}
\mathrm{d} s_{\text{global}}^2=\mathrm{d} \rho^2+\sinh^2\rho \mathrm{d} \Omega_2^2.
\end{equation}
The angular part of the metric is given by 
\begin{equation}
\mathrm{d} \Omega_2^2= -\coth^2\rho \mathrm{d} \psi^2+\mathrm{d} \phi^2.
\end{equation}
%The Laplacian is 
%\begin{equation}
%	\begin{split}
	%	\square f&=\frac{1}{\sqrt{-\text{detg}}}\partial_{\mu}\Big(\sqrt{-\text{detg}}~g^{\mu \nu}\partial_{\nu}f\Big)\\
	%	&=\frac{1}{\sqrt{-\text{detg}}}\partial_{\rho}\Big(\sqrt{-\text{detg}}~g^{\rho \rho}\partial_{\rho}f\Big)+	\csch^2\rho \square_{S^2}f\\
	%	&=\frac{1}{\sinh \rho }\partial_\rho (\sinh \rho \times  \partial_{\rho}f)+	\csch^2\rho\square_{S^2}f\\
	%	&=\frac{\partial^2f}{\partial \rho^2}+\coth \rho \frac{\partial f}{\partial \rho}+\csch^2\rho\square_{S^2}f\\
	%\end{split}
%\end{equation}
Now, we can extract the $\frac{\partial^2f}{\partial \phi^2}$ part from the angular part of the laplacian 
\begin{equation}
	\begin{split}
		\square_{S^2} f&=-\tanh^2\rho \frac{\partial^2f}{\partial \psi^2}+\frac{\partial^2f}{\partial \phi^2}\\
		\implies \frac{\partial^2f}{\partial \phi^2}&=\square_{S^2} f+\tanh^2\rho \frac{\partial^2f}{\partial \psi^2}.
	\end{split}
\end{equation}
Using it, we write the full laplacian in terms of the whole angular part, and radial part
\begin{equation}
\begin{split}
\square f&=-\sech^2\rho \frac{\partial^2f}{\partial \psi^2}+\frac{\partial^2f}{\partial \rho^2}+(\tanh \rho +\coth \rho) \frac{\partial f}{\partial \rho}+\csch^2\rho \frac{\partial^2f}{\partial \phi^2}\\
&=-\sech^2\rho \frac{\partial^2f}{\partial \psi^2}+\frac{\partial^2f}{\partial \rho^2}+(\tanh \rho +\coth \rho) \frac{\partial f}{\partial \rho}+\csch^2\rho \Bigg( \square_{S^2} f+\tanh^2\rho \frac{\partial^2f}{\partial \psi^2}\Bigg)\\
&=\frac{\partial^2f}{\partial \rho^2}+(\tanh \rho +\coth \rho) \frac{\partial f}{\partial \rho}+\csch^2\rho \square_{S^2} f.
\end{split}
\end{equation}
%\textcolor{blue}{Match with Split representation paper.}

\subsection{Sturm-Liouville equation in $\mathrm{AdS}_3 / \mathbb{Z}$}
\label{ssll}
In this subappendix \ref{ssll}, we analyze the Sturm-Liouville equation, and its solution in $\mathrm{AdS}_3 / \mathbb{Z}$. The unit metric on $\mathrm{AdS}_3 / \mathbb{Z}$ in the global coordinates is given by
\begin{equation}
\mathrm{d} s_3^2=-\cosh ^2 \rho \mathrm{d} \psi^2+\mathrm{d} \rho^2+\sinh ^2 \rho \mathrm{d} \phi^2 .
\end{equation}
The $\mathbb{Z}$ acts as the time-like quotient $\psi \sim \psi+2 \pi$. The laplacian eigenvalue equation is
\begin{equation}
\begin{split}
\square f - m^2 f &= 0 \\
\Longrightarrow \square f - \Delta(\Delta - 2) f &= 0 \\
\Longrightarrow \frac{\partial^2 f}{\partial \rho^2} + (\tanh \rho + \operatorname{coth} \rho) \frac{\partial f}{\partial \rho} + \operatorname{csch}^2 \rho \square_{S^2} f &= \Delta(\Delta - 2)f \\
\Longrightarrow \frac{\partial^2 f}{\partial \rho^2} + 2 \operatorname{coth} 2\rho \,\frac{\partial f}{\partial \rho} + \operatorname{csch}^2 \rho \square_{S^2} f &= \Delta(\Delta - 2)f.
\end{split}
\end{equation}

Little group at $X_0=(1,0,\cdots,0)$ is $H_{X_0}=\mathrm{SO}(2,1)$. If $f$ is $H_{X_0}$-left-invariant function on $\mathrm{AdS}_3/\mathbb{Z}$,
the radial part gives the following Sturm-Liouville equation, 
\begin{equation}
\begin{aligned}
    \frac{\partial^2 f}{\partial \rho^2}+2 \operatorname{coth} 2\rho \,\frac{\partial f}{\partial \rho} &=\Delta(\Delta-2)f.
\end{aligned}
\end{equation}
If we set $\operatorname{cosh}2\rho = z$, we can write the above equation as,
\begin{equation}
\begin{split}
    (1-z^2) \frac{\partial^2 f}{\partial z^2} - 2 z \frac{\partial f}{\partial z} + \frac{1}{4}\Delta(\Delta-2) f &= 0.
\end{split}
\end{equation}
We can recognize it as the Gegenbauer Polynomial equation
\begin{equation}
\left(1-x^2\right) y^{\prime \prime} - (2 \alpha + 1) x y^{\prime} + n(n+2 \alpha) y = 0 .
\end{equation}
We make a table for illustration: $\alpha=0$, $\alpha=1 / 2$, and $\alpha=1$.
\begin{table}[h]
\centering

\begin{tabular}{|l|l|l|}
\hline
Case & Differential Equation & Polynomials \\
\hline
$\alpha = 0$ &  Chebyshev differential equation & Chebyshev polynomials of the first kind \\
\hline
$\alpha = 1/2$ & Legendre equation & Legendre polynomials \\
\hline
$\alpha = 1$ & Chebyshev differential equation & Chebyshev polynomials of the second kind \\
\hline
\end{tabular}
\caption{Special Cases of Gegenbauer Polynomials}
\end{table}

%\begin{itemize}
%\item When $\alpha=1 / 2$, the equation reduces to the Legendre equation, and the Gegenbauer polynomials reduce to the Legendre polynomials.\\
%\item When $\alpha=1$, the equation reduces to the Chebyshev differential equation, and the Gegenbauer polynomials reduce to the Chebyshev polynomials of the second kind.
%\end{itemize}
%\end

The solution to the differential equation are given as Gaussian hypergeometric series in certain cases where the series is in fact finite
\begin{equation}
C_n^{(\alpha)}(z) = \frac{(2 \alpha)_n}{n!} {}_2 F_1\left(-n, 2 \alpha+n ; \alpha+\frac{1}{2} ; \frac{1-z}{2}\right) .
\end{equation}

Here $(2 \alpha)_n$ is the rising factorial. Explicitly,
\begin{equation}
C_n^{(\alpha)}(z) = \sum_{k=0}^{\lfloor n / 2 \rfloor} (-1)^k \frac{\Gamma(n-k+\alpha)}{\Gamma(\alpha) k !(n-2 k) !} (2 z)^{n-2 k} .
\end{equation}
From this it is also easy to obtain the value at unit argument
\begin{equation}
C_n^{(\alpha)}(1)=\frac{\Gamma(2 \alpha+n)}{\Gamma(2 \alpha) n !} .
\end{equation}
They are special cases of the Jacobi polynomials
\begin{equation}
C_n^{(\alpha)}(x)=\frac{(2 \alpha)_n}{(\alpha+\frac{1}{2})_n} P_n^{(\alpha-1 / 2, \alpha-1 / 2)}(x) .
\end{equation}
in which $(\theta)_n$ represents the rising factorial of $\theta$.
One therefore also has the Rodrigues formula
\begin{equation}
C_n^{(\alpha)}(x)=\frac{(-1)^n}{2^n n !} \frac{\Gamma(\alpha+\frac{1}{2}) \Gamma(n+2 \alpha)}{\Gamma(2 \alpha) \Gamma(\alpha+n+\frac{1}{2})}\left(1-x^2\right)^{-\alpha+1 / 2} \frac{\mathrm{d}^n}{\mathrm{d} x^n}\left[\left(1-x^2\right)^{n+\alpha-1 / 2}\right].
\end{equation}
\paragraph{Solution to the differential equation for the $\mathrm{AdS}_3/\mathbb{Z}$.}
Our differential equation for the $\mathrm{AdS}_3/\mathbb{Z}$ belongs to the $\alpha=\frac{1}{2}$ case, i.e. they are Legendre polynomials with $n= -\frac{\Delta}{2}$ or $n=\frac{(\Delta-2)}{2}=\frac{\Delta}{2}-1$. The solution to the differential equation can be written as
\begin{equation}
C_{-\frac{\Delta}{2}}^{\frac{1}{2}}(z)={ }_2 F_1\left(\frac{\Delta}{2}, 1-\frac{\Delta}{2} ; 1 ; \frac{1-z}{2}\right)= P_{-\frac{\Delta}{2}}(z) ,
\end{equation}
\begin{equation}
C_{\frac{\Delta}{2}-1}^{\frac{1}{2}}(z)={ }_2 F_1\left(1-\frac{\Delta}{2}, \frac{\Delta}{2} ; 1 ; \frac{1-z}{2}\right)= P_{\frac{\Delta}{2}-1}(z) .
\end{equation}
The relations above are known from Murphy's Formula for Legendre Polynomials.

\subsection{Remark on the dependence of dimension: in arbitrary dimension}
\label{aadd}
In this subappendix \ref{aadd}, we give remark on the dependence of dimension: in arbitrary dimension.
Now, in arbitrary dimension
\begin{equation}
	\mathrm{d} s_{\text{AdS}_{d+1}}^2=\mathrm{d} \rho^2+\sinh^2\rho \mathrm{d} \Omega_d^2.
\end{equation}
Now, in another form we have  
\begin{equation}
	\mathrm{d} s_{\text{AdS}_{d+1}}^2= \mathrm{d} \rho^2+\cosh^2\rho  \mathrm{d} \psi^2+\sinh^2\rho \mathrm{d} \Omega_{d-1}^2.
\end{equation}
We get the $\mathrm{d}\Omega_d^2$ 
\begin{equation}
\begin{split}
\sinh^2\rho \mathrm{d} \Omega_d^2&=\cosh^2\rho  \mathrm{d} \psi^2+\sinh^2\rho \mathrm{d} \Omega_{d-1}^2\\
\implies  \mathrm{d}\Omega_d^2&=\coth^2\rho  \mathrm{d} \psi^2+ \mathrm{d} \Omega_{d-1}^2.
\end{split}
\end{equation}
The laplacian is 
\begin{equation}
	\begin{split}
		\square f
  %&=\frac{1}{\sqrt{\text{detg}}}\partial_{\mu}\Big(\sqrt{\text{detg}}~g^{\mu \nu}\partial_{\nu}f\Big)\\
		%&=\frac{1}{\sqrt{\text{detg}}}\partial_{\psi}\Big(\sqrt{\text{detg}}~g^{\psi \psi}\partial_{\psi}f\Big)+\frac{1}{\sqrt{\text{detg}}}\partial_{\rho}\Big(\sqrt{\text{detg}}~g^{\rho \rho}\partial_{\rho}f\Big)+\frac{1}{\sqrt{\text{detg}}}\partial_{S^{d-1}}\Big(\sqrt{\text{detg}}~g^{\Omega \Omega }\partial_{S^{d-1}}f\Big)\\
		%&=\frac{1}{\sinh^{d-1}\rho \cosh \rho}\partial_\psi (\sinh^{d-1} \rho \cosh \rho \times \sech^2\rho \partial_{\psi}f)+\frac{1}{\sinh^{d-1} \rho \cosh \rho}\partial_\rho (\sinh^{d-1} \rho \cosh \rho  \partial_{\rho}f)\\
		%&+\frac{1}{\sinh^{d-1} \rho \cosh \rho}\partial_{S^{d-1}} (\sinh^{d-1} \rho \cosh \rho \times \csch^2\rho \partial_{S^{d-1}}f)\\
		&=\sech^2\rho \frac{\partial^2f}{\partial \psi^2}+\frac{\partial^2f}{\partial \rho^2}+(\tanh \rho +(d-1)\coth \rho) \frac{\partial f}{\partial \rho}+\csch^2\rho \square_{S^{d-1}}f.
	\end{split}
\end{equation}
We can get $\square_{S^{d-1}}f$
\begin{equation}
	\begin{split}
	\square_{S^d} f&=\tanh^2\rho \frac{\partial^2f}{\partial \psi^2}+\square_{S^{d-1}}f\\
	\implies \square_{S^{d-1}}f &=\square_{S^d} f-\tanh^2\rho \frac{\partial^2f}{\partial \psi^2}.
\end{split}
\end{equation}
Therefore, we have 
\begin{equation}
	\begin{split}
		\square f&=\sech^2\rho \frac{\partial^2f}{\partial \psi^2}+\frac{\partial^2f}{\partial \rho^2}+(\tanh \rho +(d-1)\coth \rho) \frac{\partial f}{\partial \rho}+\csch^2\rho \square_{S^{d-1}}f\\
		&=\sech^2\rho \frac{\partial^2f}{\partial \psi^2}+\frac{\partial^2f}{\partial \rho^2}+(\tanh \rho +(d-1)\coth \rho) \frac{\partial f}{\partial \rho}+\csch^2\rho \Bigg(\square_{S^d} f-\tanh^2\rho \frac{\partial^2f}{\partial \psi^2}\Bigg)\\
		&=\frac{\partial^2f}{\partial \rho^2}+(\tanh \rho +(d-1)\coth \rho) \frac{\partial f}{\partial \rho}+\csch^2\rho \square_{S^d} f.
	\end{split}
\end{equation}
%From which we will get 
%\begin{equation}
%\frac{\partial}{\partial \rho}\Bigg(e^{2\rho}\frac{\partial f}{\partial \rho}\Bigg)=\Delta (\Delta-2)e^{2\rho}f.
%\end{equation}

%\begin{equation}
%\frac{\partial}{\partial \rho}\Bigg(e^{-2\rho}\frac{\partial f}{\partial \rho}\Bigg)=\Delta (\Delta-2)e^{-2\rho}f.
%\end{equation}
\section{Alternative derivation of the Plancherel measure as a special case: case $p=q=2$}
\label{alder}
In this appendix \ref{alder}, we give an alternative derivation of the Plancherel measure as a special case: case $p=q=2$.

First, we compute some necessary quantities to compute the ingredients for the Plancherel measure, i.e. $a(\rho)$, and $b(\rho)$.
\paragraph{Case $p=q=2$.}

First, we compute 
\begin{equation}
\mathfrak{g}(\rho,0,0,2,2)=4\int_{0}^{\pi} \int_{0}^{\pi} |\cos \theta_1-\cos\theta_2|^{-1+i\rho} \mathrm{d}\theta_1 \mathrm{d}\theta_2,
\end{equation}
and 
\begin{equation}
\mathfrak{g}(\rho,1,0,2,2)=4\int_{0}^{\pi} \int_{0}^{\pi} |\cos \theta_1-\cos\theta_2|^{-1+i\rho} \textrm{sgn}(\cos\theta_1-\cos\theta_2)\cos\theta_1\mathrm{d}\theta_1 \mathrm{d}\theta_2.
\end{equation}
We use that
\begin{equation}
\cos \theta_1 - \cos \theta_2 = -2 \sin \left(\frac{\theta_1 - \theta_2}{2}\right) \sin \left(\frac{\theta_1 + \theta_2}{2}\right).
\end{equation}
We make change of variables 
\begin{equation}
\begin{split}
 r = \frac{\theta_1 - \theta_2}{2}~~,~~ s = \frac{\theta_1 + \theta_2}{2}.
\end{split}
\end{equation}
Then, we have
\begin{equation}
\begin{split}
\mathfrak{g}(\rho,0,0,2,2)&=2^{2+i\rho}\times  \int_{0}^{\frac{\pi}{2}}\left( \int_{-s}^{s}|\sin r|^{-1+i\rho}  \mathrm{d}r\right)|\sin s|^{-1+i\rho} \mathrm{d}s\\
&+2^{2+i\rho} \times \int_{\frac{\pi}{2}}^{\pi}\left( \int_{s-\pi}^{\pi-s}|\sin r|^{-1+i\rho}  \mathrm{d}r\right)|\sin s|^{-1+i\rho} \mathrm{d}s\\
&=2^{4+i\rho}\times  \int_{0}^{\frac{\pi}{2}} \left( \int_{0}^{s}|\sin r|^{-1+i\rho}  \mathrm{d}r\right)|\sin s|^{-1+i\rho} \mathrm{d}s.
\end{split}
\end{equation}

\begin{equation}
\begin{split}
\mathfrak{g}(\rho,1,0,2,2)&=2^{2+i\rho}\times  \int_{0}^{\frac{\pi}{2}}\left( \int_{-s}^{s}|\sin r|^{-1+i\rho} \cos(r+s)\textrm{sgn}r  \mathrm{d}r\right)|\sin s|^{-1+i\rho} \mathrm{d}s\\
&+2^{2+i\rho} \times \int_{\frac{\pi}{2}}^{\pi}\left( \int_{s-\pi}^{\pi-s}|\sin r|^{-1+i\rho}\cos(r+s)\textrm{sgn}r  \mathrm{d}r\right)|\sin s|^{-1+i\rho} \mathrm{d}s\\
&=-2^{4+i\rho}\times  \int_{0}^{\frac{\pi}{2}} \left( \int_{0}^{s}|\sin r|^{i\rho}  \mathrm{d}r\right)|\sin s|^{i\rho} \mathrm{d}s.
\end{split}
\end{equation}
Here, we use 
\begin{equation}
\cos(r+s)=\cos r \cos s-\sin r \sin s,
\end{equation}
and the term $\cos r \times \text{sgn}r |\sin r|^{-1+i\rho}$ integrates to zero because it is odd. 
Now, we evaluate the integrals. 
%Using the substitution 
%\begin{equation}
%\sin \theta =\sqrt{1-u}
%\end{equation}
Now, we have
\begin{equation}
\begin{split}
\int_{0}^{\frac{\pi}{2}} \left( \sin \theta\right)^{\alpha}\mathrm{d} \theta&=\frac{1}{2}\beta\left(\frac{\alpha+1}{2},\frac{1}{2} \right)\\
&=\frac{\sqrt{\pi}}{2}\frac{\Gamma\left(\frac{\alpha+1}{2}\right)}{\Gamma(\alpha+1)}.
\end{split}
\end{equation}
%Writing 
%\begin{equation}
%\varphi_{\alpha}(s)=\int_{0}^{s}%%(\sin\theta)^{\alpha}\mathrm{d}\theta
%\end{equation}
%we find
We find 
\begin{equation}
\int_{0}^{\frac{\pi}{2}} \left( \int_{0}^{s}(\sin r)^{\alpha}  \mathrm{d}r\right)(\sin s)^{\alpha} \mathrm{d}s=\frac{\pi}{8}\frac{\Gamma\left(\frac{\alpha+1}{2}\right)^2}{\Gamma(\alpha+1)}.
\end{equation}
We have
\begin{equation}
\begin{split}
\mathfrak{g}(\rho,0,0,2,2)&=2^{1+i\rho}\pi \frac{\Gamma(i\rho/2)^2}{\Gamma\left( 1/2+i\rho/2\right)^2}\\
\mathfrak{g}(\rho,1,0,2,2)&=-2^{1+i\rho}\pi \frac{\Gamma(1/2+i\rho/2)^2}{\Gamma\left( 1+i\rho/2\right)^2}.
\end{split}
\end{equation}
We find
\begin{equation}\label{arhobrho}
\begin{split}
a(\rho)&=\frac{1}{4\pi}\left|\rho \tanh\frac{\pi\rho}{2}\right|\\
b(\rho)&=\frac{1}{4\pi}\left|\rho \coth\frac{\pi\rho}{2}\right|.
\end{split}
\end{equation}
The measures $a(\rho)^2 \mathrm{d} \rho$ and $b(\rho)^2 d \rho$ represent the Plancherel measures.

\providecommand{\href}[2]{#2}\begingroup\raggedright\endgroup
	
\end{document}